\documentclass[iop]{emulateapj}
\usepackage{natbib}
\usepackage{longtable, tabu}
\usepackage{tablefootnote}
\usepackage{footnote}
\usepackage{tabularx}
\usepackage{graphicx}
\usepackage{amsmath}
\usepackage{subfigure}
\usepackage{lipsum}
\usepackage{float}
\usepackage{color}
\usepackage{array,booktabs}
\usepackage[toc,page]{appendix}

\newcounter{ltfootnote}



\newcolumntype{n}{>{\centering\arraybackslash}X}

\def\kepler{{\slshape Kepler}}
\def\K2{{\slshape K2}}
\def\spitzer{{\it Spitzer}}
\def\HST{{\slshape HST}}
\def\JWST{{\slshape JWST}}

\newcommand{\myemail}{shryguo@mit.edu}

\slugcomment{submitted to ApJ on 03/30/2018}

\begin{document}
\title{Temperate super-Earths/mini-Neptunes around M/K dwarfs Consist of 2 Populations Distinguished by Their Atmospheres}

\author{Xueying~Guo\altaffilmark{1, a}}

\author{Sarah~Ballard\altaffilmark{1}}

\author{Diana~Dragomir\altaffilmark{1, b}}

\author{Michael~Werner\altaffilmark{2}}

\author{Varoujan~Gorjian\altaffilmark{2}}

\altaffiltext{A}{\myemail}

\altaffiltext{1}{MIT Kavli Institute for Astrophysics \& Space Research, Cambridge, MA 02139, USA}

\altaffiltext{2}{Jet Propulsion Laboratory, 4800 Oak Grove Drive, Pasadena, CA, USA}

\altaffiltext{B}{NASA Hubble Fellow}


\begin{abstract}

Studies of the atmospheres of hot Jupiters reveal a diversity of atmospheric composition and haze properties. Similar studies on individual smaller, temperate planets are rare due to the inherent difficulty of the observations and also to the average faintness of their host stars. To investigate their ensemble atmospheric properties, we construct a sample of 28 similar planets, all possess equilibrium temperature within 300--500~K, have similar size (1--3~$R_{\oplus}$), and orbit early M dwarfs and late K dwarfs with effective temperatures within a few hundred Kelvin of one another. In addition, NASA's \kepler/\K2\ and \spitzer\ missions gathered transit observations of each planet, producing an uniform transit data set both in wavelength and coarse planetary type. With the transits measured in \kepler's broad optical bandpass and \spitzer's 4.5 $\mu$m wavelength bandpass, we measure the transmission spectral slope, $\alpha$, for the entire sample. While this measurement is too uncertain in nearly all cases to infer the properties of any individual planet, the distribution of $\alpha$ among several dozen similar planets encodes a key trend. We find that the distribution of $\alpha$ is not well-described by a single Gaussian distribution. Rather, a ratio of the Bayesian evidences between the likeliest 1-component and 2-component Gaussian models favors the latter by a ratio of 100:1. One Gaussian is centered around an average $\alpha = -1.3$, indicating hazy/cloudy atmospheres or bare cores with atmosphere evaporated. A smaller but significant second population ($20\pm 10\%$ of all) is necessary to model significantly higher $\alpha$ values, which indicate atmospheres with potentially detectable molecular features. We conclude that the atmospheres of small and temperate planets are far from uniformly flat, and that a subset are particularly favorable for follow-up observation from space-based platforms like the Hubble Space Telescope and the James Webb Space Telescope.

\end{abstract}

\bigskip

\section{Introduction}

Astronomers are beginning to uncover the diversity of planet atmospheres. While studies of the atmospheres of Hot Jupiters number in the dozens \citep{Sing2016}, there are to date only 4 planets smaller than 3 $R_{\oplus}$ with published transmission spectra. Two of these planets, GJ~1214b \citep{Charbonneau2009} and GJ~1132b \citep{Berta2015}, orbit nearby M dwarfs. An extensive campaign from the ground \citep{Bean2010}, with the Spitzer Space Telescope \citep{Desert2011, Fraine2013}, and with the Hubble Space Telescope \citep{Berta2012, Kreidberg2014} revealed a flat and featureless transmission spectrum for GJ~1214b. For GJ~1132 in contrast, a deeper transit in z band may indicate strong opacity from $\rm H_2 O$ or $\rm CH_4$ \citep{Southworth2017}. A third and much hotter planet, 55~Cnc~e, transits a nearby G~dwarf \citep{winn2011, Demory2011}, and the fourth planet, HD~97658b, transits a K~dwarf with the equilibrium temperature of around 750~K \citep{VanGrootel2014, Dragomir2013, Knutson2014}. The paucity of transmission spectra for small planets, despite their common occurrence in the Milky Way \citep{Howard2012, Dressing2013, Dressing2015, Fressin2013}, is due to the faintness of most of their host stars, uncovered by NASA's \kepler\ mission. Additional studies of the individual transmission spectra of small planets await their discoveries around bright and nearby stars. NASA's {\it TESS} Mission, to be launched in the Spring of 2018, will uncover hundreds of planets smaller than 3 $R_{\Earth}$, with dozens orbiting stars brighter than 10th magnitude at optical wavelengths \citep{Sullivan2015, Muirhead2017, Ballard2018}.


While the individual transmission spectra of small planets are largely the purview of studies a few years in the future, we already have in hand sufficient data to investigate at least one average property of the atmospheres of sub-Neptune planets. There exist more than two dozen small (1--3~$\rm R_{\oplus}$) and temperate (300--500~K) exoplanets, all orbiting M or late K dwarfs, with transit observations observed from both NASA's \kepler/\K2\ mission (with a broad optical bandpass) and NASA's \spitzer\ space telescope (at 4.5~$\mu$m). A difference in transit depth for a given planet between these bandpasses, while far from conclusive, encodes useful information about the planetary atmosphere. 

Recently, \cite{Sing2016} found that the difference in transit depth between the blue-optical and mid-infrared is a useful discriminant in Jovian atmospheres. They measured the difference between the optical and near-infrared transit depths in units of scale height, and identified a strong correlation between this quantity with the strength of the water absorption feature at 1.4~$\mu$m. For little-to-no difference between the transit depths at optical and mid-infrared wavelengths, the water feature was stronger, whereas a negative spectral slope (deeper optical transit than near-infrared transit) was associated with weak or no water absorption. Crucially for M~dwarfs, a significant wavelength dependence in transit depth implies that an atmosphere of some kind exists. This is a critical constraint when the retention of any atmosphere at all for longer than a Gyr on a small planet orbiting an M~dwarf is uncertain due to stellar activity, tidal interactions, and other effects \citep{Zendejas2010, Barnes2013, Heng2012, Shields2016}. Nature creates radically different atmospheric conditions even within much similar temperature and radius ranges, as Earth and Venus illustrate. With this sample, rather than characterize any one planet, we aim to investigate the average of the group as a whole, or indeed to establish whether the sample can be well-described by a single normal distribution. While the Spitzer transit precision may be insufficient to distinguish a flat transmission spectrum from one with features on a case-by-case basis, a joint analysis of two dozen similar planets enables an average measurement on the transmission spectra of planets of this kind.

A few reasons can lead to the difference between the \kepler\ band and the \spitzer\ band transit depths of a planet. In a blended binary or a blended authentic planetary system, the transit depth will vary unless the two stars have identical effective temperatures \citep{Fressin2010}. As the planet blocks a fraction of one star, the budget of combined photons from the two stars will shift redward or blueward, producing an apparent change to the transit depth with wavelength. In this work, we will carefully select our targets so that, in combination with \spitzer's high spatial resolution, the effect of a blended nearby source on our \spitzer\ transit depth measurements will be negligible. And we correct for the blended source effect on the \kepler\ transit depth measurements with the radius correction factors from \cite{Furlan2017}. Alternatively, star spots can also produce an artificially different transit depth between two wavelengths \citep{McCullough2014}, with strongest effect on transit depth in the blue. In this case, the near-contemporaneous nature of the \kepler\ and \spitzer\ data sets for each planet is key: we have an estimation for how the star spot coverage is varying on the star from the optical light curve. The remaining explanations for a different transit depth are atmospheric. The observed transit depth changes because the optical depth of the atmosphere varies with wavelength and alters the size of the planet’s silhouette. We adopt a phenomenological approach rather than a physical one for the transmission spectral slope: an exhaustive list of the variety of atmospheric properties that could produce a different transit depth between these two bandpasses is beyond the scope of this paper. There exist three possibilities for each planet: the transit depth may be indistinguishable between the two bandpasses, it may be deeper in the infrared (in which case, molecular absorption must play a role of some kind), or it maybe be shallower in the infrared (in which case, we assume haze-like scattering is occurring, either with or without molecular absorption). The \kepler\ bandpass and the \spitzer\ 4.5~$\mu$m bandpass are sufficiently far apart to probe the average transmission spectral slope \citep{Sing2016}, and investigate, albeit coarsely, the general trend of the transmission spectra of these planets. 

The layout of this paper is as follows. In Section \ref{sec:sample}, we describe our our sample planets and their \spitzer\ observation parameters. In Section \ref{sec:fitting}, we describe the Pixel-Level Decorrelation and Gaussian process that we used to analyze the light curves, and the transit parameter selections. The transit fitting results and the comparison with previous works are presented in Section \ref{sec:results}. And we discuss in detail the atmosphere models we use to calculate the transmission spectra slopes and the cross section power law index in Section \ref{sec:atmosphere}. The ensemble atmospheric properties of our planet sample is also presented in this section, along with their trend as a function of different planetary properties. We present more considerations into the calculation of scale heights of planet atmospheres in the discussion of Section \ref{sec:discussion}, as well as future directions into improving the study of atmospheres of small temperate planets.

\bigskip

\section{Planet Sample and Data}\label{sec:sample}

Our sample comprises planets with equilibrium temperatures between 300--600~K and radii between 1--3~$R_{\oplus}$, orbiting stars of spectral type M or late K. All stars within this sample possess published spectroscopic temperatures and radii \citep{Mann2013, Martinez2017, Crossfield2016, Dressing2017, Huber2016}. We include only stars with spectroscopic characterization for several reasons: first, we can be certain that the host stars are dwarfs and not giants. Secondly, with a typical Kepler M~dwarf spectrum and exoplanet light curve in hand, the planetary radius and insolation are known to 10\% and 20\% certainty \citep{Muirhead2012, Ballard2013}, while these errors are 2-3 times larger for planets transiting Kepler M~dwarfs with only photometric stellar characterization \citep{Dressing2013}. Of the approximately 490 hours of data for these 28 planets, amounting to 64 total transits, approximately half are published in the literature \citep{Desert2015, Ballard2011}. Of our 28 planets, 18 possess effective temperatures in the approximate M1V-K7V range between 3800--4200~K \citep{Boyajian2013}. An additional 9 reside within the slightly cooler range 3400--3800~K, bracketing spectral type M3V. Additionally in this sample, all \spitzer\ observations were gathered either contemporaneous with \kepler\ observations of the planet or within a year of them. The properties of our sample planets and their host stars extracted from previous works, and the observation channels and duration are listed in Table \ref{tab:sample}.

In our sample of 28 planets, 18 are from the original \kepler\ mission (\spitzer\ Program ID 60028 with PI David Charbonneau) and 10 are \K2\ planets (\spitzer\ Program ID 11026 with PI Michael Werner). All 28 planets were observed by the \spitzer\ Infrared Array Camera (hereafter IRAC) at 4.5~$\mu$m, while 2 of them were also observed by the IRAC at 3.6~$\mu$m. Each observation of a planet covers one transit, with the light curve cadence ranging between 2 seconds and 30 seconds. 
Physical properties of planets in our sample and their host stars are extracted from a set of previous works with the most accurate and precise results, which are described in detail in section \ref{sec:SetParameters}. The number of archived \spitzer\ observations of each planet that we used is also presented in Table \ref{tab:sample}.

\begin{table*}
\renewcommand*{\arraystretch}{1.7}
    \caption{Properties of Small, Cool {\it Kepler}/K2 M Dwarf Planets Observed by \spitzer\ }
    \centering
    \begin{tabularx}{\textwidth}{cnnnnnnnnnn}
    \hline\hline
    Name & $N_{\rm{tranet}}$\footnote{Number of transit planets in the system.} & $K_s$ (mag) & $T_{\rm{eff}}$\footnote{The effective temperature of the host star.} (\rm K) & $R_{\rm p}$ ($R_{\oplus}$) & P(days) & $T_{\rm{p}}$ (\rm K) & $N_{Spitzer}$\footnote{Number of transits observed by \spitzer.} & IRAC Channel & obs. time (hrs) \\ 
    \hline
    KOI247.01 & 1 & 11.1 & 3852$\pm$60 & 1.95 & 13.81 & 391 & 3 & 4.5 $\mu$m & 4.86 \\ 
    Kepler-49b & 4 & 12.4 & 3970$\pm$60 & 2.63 & 7.20 & 509 & 2 & 4.5 $\mu$m & 6.50 \\ 
    Kepler-49c & 4 & 12.4 & 3970$\pm$60 & 2.13 & 10.91 & 443 & 3 & 4.5 $\mu$m & 5.61 \\ 
    Kepler-504b & 1 & 11.2 & 3548$\pm$60 & 1.59 & 9.55 & 374 & 2 & 4.5 $\mu$m & 5.14 \\ 
    Kepler-26c & 4 & 12.6 & 4049$\pm$60 & 2.72 & 17.25 & 385 & 3 & 4.5 $\mu$m & 6.16 \\ 
    Kepler-125b & 2 & 11.7 & 3770$\pm$60 & 2.67 & 4.16 & 590 & 2 & 4.5 $\mu$m & 5.26 \\ 
    KOI252.01 & 1 & 12.6 & 3858$\pm$60 & 2.64 & 17.60 & 362 & 2 & 4.5 $\mu$m & 8.67 \\ 
    KOI253.01 & 2 & 12.3 & 3900$\pm$60 & 2.73 & 6.38 & 530 & 1 & 4.5 $\mu$m & 6.16 \\ 
    Kepler-505b & 3 & 12.1 & 3900$\pm$60 & 2.60 & 27.52 & 314 & 1 & 4.5 $\mu$m & 9.69 \\ 
    Kepler-138c & 3 & 9.51 & 3871$\pm$60 & 1.20 & 13.78 & 402 & 1 & 4.5 $\mu$m & 5.42 \\ 
    Kepler-138d & 3 & 9.51 & 3871$\pm$60 & 1.21 & 23.09 & 338 & 4 & 4.5 $\mu$m & 8.09 \\ 
    Kepler-205c & 2 & 10.8 & 4121$\pm$60 & 1.58 & 20.31 & 389 & 2 & 4.5 $\mu$m & 7.62 \\ 
    Kepler-236c & 2 & 12.3 & 3866$\pm$60 & 2.00 & 23.97 & 329 & 2 & 4.5 $\mu$m & 8.23 \\ 
    Kepler-249d & 3 & 12.0 & 3630$\pm$60 & 1.08 & 15.37 & 332 & 2 & 4.5 $\mu$m & 6.80 \\ 
    Kepler-737b & 1 & 12.1 & 3792$\pm$60 & 1.96 & 28.60 & 298 & 2 & 4.5 $\mu$m & 9.26 \\ 
    Kepler-32d & 5 & 12.8 & 3801$\pm$60 & 2.50 & 22.78 & 328 & 2 & 4.5 $\mu$m & 8.15 \\ 
    Kepler-36b & 1 & 12.3 & 4158$\pm$60 & 2.24 & 59.88 & 273 & 1 & 4.5 $\mu$m & 11.51 \\ 
    Kepler-395c & 2 & 12.9 & 3915$\pm$60 & 1.34 & 34.99 & 292 & 3 & 4.5 $\mu$m & 11.61 \\ 
    K2-3b & 3 & 8.56 & 3976$\pm$205 & 2.18 & 10.05 & 463 & 6 & 4.5 $\mu$m & 6.40 \\ 
    K2-3c & 3 & 8.56 & 3976$\pm$205 & 1.85 & 24.64 & 344 & 3 & 4.5 $\mu$m & 7.26 \\ 
    K2-3d & 3 & 8.56 & 3976$\pm$205 & 1.51 & 44.56 & 282 & 2 & 4.5 $\mu$m & 7.84 \\ 
    K2-9b & 1 & 11.5 & 3460$\pm$164 & 2.25 & 18.45 & 314 & 2 & 4.5 $\mu$m & 8.54 \\ 
    K2-18b & 1 & 8.90 & 3527$\pm$162 & 2.38 & 32.94 & 272 & 1 & 4.5 $\mu$m & 7.65 \\ 
    K2-21b & 2 & 9.42 & 3952$\pm$202 & 1.84 & 9.32 & 500 & 1 & 4.5 $\mu$m & 14.81 \\ 
    K2-21c & 2 & 9.42 & 3952$\pm$202 & 2.49 & 15.50 & 420 & 1 & 4.5 $\mu$m & 14.81 \\ 
    K2-26b & 1 & 10.5 & 3769$\pm$57 & 2.67 & 14.57 & 430 & 1 & 4.5 $\mu$m & 6.73 \\ 
    K2-28b & 1 & 10.7 & 3293$\pm$88 & 2.32 & 2.26 & 568 & 1 & 4.5 $\mu$m & 7.98 \\ 
    EPIC 210558622 & 1 & 9.50 & 4581$\pm$149 & 3.20 & 19.56 & 490 & 2 & 4.5 $\mu$m & 7.98 \\ 
    \hline
        Kepler-138d & 3 & 9.51 & 3871$\pm$60 & 1.21 & 23.09 & 338 & 4 & 3.6 $\mu$m & 8.09 \\ 
        K2-18b & 1 & 8.90 & 3527$\pm$162 & 2.38 & 32.94 & 272 & 1 & 3.6 $\mu$m & 7.98 \\ 
    \hline
    \end{tabularx}
    \label{tab:sample}
\bigskip
\end{table*}

\bigskip

\section{Spitzer Light Curve Fitting}\label{sec:fitting}

We model our \spitzer\ light curves using the Pixel-Level Decorrelation (hereafter PLD) method, as well as a Gaussian process. as described in \cite{Deming2015}, which is the most effective way of dealing with the flux fluctuations of \spitzer\ IRAC images that are caused by pointing jitter. This methodology, which we describe in detail in this section, is applicable under two circumstances. First, PLD relies upon the assumption that the incoming stellar flux is falling on the same set of pixels throughout the entire time series. For a constant source, the values of these pixels ought to sum to the same amount at every exposure. When it does not, as for both \spitzer\ \citep{Deming2015} and \kepler\ observations \citep{Luger2016, Luger2017}, then we know pixel sensitivity variations must be responsible. The under-sampled PSF may drift only a fraction of a single pixel over the duration of several hours, but sensitivity changes over the pixel will cause a constant astrophysical source to appear as much as 10\% brighter or dimmer. This effect, which swamps the hundreds-of-ppm astrophysical signal we hope to extract, requires careful treatment to remove. Secondly, while \cite{Deming2015} demonstrated that PLD can achieve shot-limited precision, a Gaussian process used in tandem can characterize any residual systematic noise (whether due to instrumental or astrophysical signal). The use of Gaussian processes for the treatment of systematic noise relies upon the selection of the correct kernel to characterize the correlation timescale and amplitude. While timescale and amplitude are free ``hyper parameter'' that are allowed to vary, the functional form itself of the kernel must be chosen judiciously. In recent years, the number of total observations $N$ was also a limiting factor for the use of Gaussian processes, which requires an $N\times N$ matrix inversion \citep{Gibson2012}. However, recent innovations have reduced this computing time from O($N^2$) to O($N$) \citep{Foreman-Mackey2017}.

The fitting parameters are composed of nuisance parameters describing the weight of each pixel included and systematic temporal ramps, and transit model parameters. On top of that, we cooperate a Gaussian process into the fitting to account for the possible red noise in \spitzer\ time series. The covariance matrix of the Gaussian process is parameterized with a Matern$-3/2$ kernel, which has been widely used in previous light curve analyses and proved to perform well \citep{Louden2017, Gibson2013}. This kernel is defined by two hyper-parameters: $\rho$, which encodes the autocorrelation timescale, and amplitude $A$, which encodes the fractional contribution from off-diagonal elements to the total covariance. For A = 0, the noise is purely Gaussian.

\subsection{Pixel-Level Decorrelation}


We employ the PLD method described in \citep{Deming2015}, which is the most effective way of dealing with the flux fluctuations of \spitzer\ IRAC images that are caused by pointing jitters, and we briefly summarize as follows. The PSF moves by less than a tenth of one pixel during the entire time series \citep{Grillmair2012}, so that the same set of typically 9 pixels encodes both the incident flux and the relative position of the star. We illustrate the PLD process first with an example of a constant astrophysical source. As pointing drift occurs, the relative flux received by each pixel changes slightly, but the total flux incident upon the pixels remains the same. At every point in time, we know a priori that the 9 pixels must sum to the same constant number. To enforce this sensible conclusion, we normalize the set of 9 pixel values at each exposure time. We call these $P^{t}$. The linear combination of these 9 individual pixel time series is the measured time series: in an ideal scenario, we would weigh each pixel uniformly in this sum. However, if the pixels are not perfectly uniform, we ought to weigh each pixel light curve $P_i^{t}$ according to this variation, where i runs from 1 to 9. Mathematically, we treat each individual normalized pixel’s time series $P_i^{t}$ as a eigenvector, where the eigenvalues encode the relative pixel sensitivities. It remains only to solve for the 9 eigenvalues, $c_i$. We add only two additional eigenvectors to describe temporal variation in the astrophysical source: the transit model, and a temporal ramp. Equation \ref{PLD} shows the formula of the PLD model at each time $t$ (details of the PLD algorithm could be seen in Section 2 of \cite{Deming2015}):

\begin{equation*}\label{PLD}
\Delta S^t = \sum_{i=1}^{N} c_i \hat{P}_i^t + DE(t) + ft + gt^2 + h,
\end{equation*}

\noindent
where $\Delta S^t$ is the variation of total brightness at time $t$, $\hat{P}_i^t$ is the relative flux change contributed by pixel $i$, which is scaled by a weight parameter $c_i$, $D$ is the transit depth and $E(t)$ is the normalized transit shape, $ft + gt^2 + h$ is the polynomial to describe the detector temporal ramp. In practice, we find that the effect of the constant term $h$ can be included in the variation of the set of weight parameters $\{c_i\}$, so our physical model has $N+3$ free parameters in total, where $N$ is the pixel number. In section \ref{sec:GP}, we discuss the Gaussian Process (GP hereafter) that is introduced in our fitting process to account for the time-correlated noise, and this adds in 2 more parameters, making the total number of parameters $N+5$.

\begin{figure}
 \includegraphics[width=\linewidth]{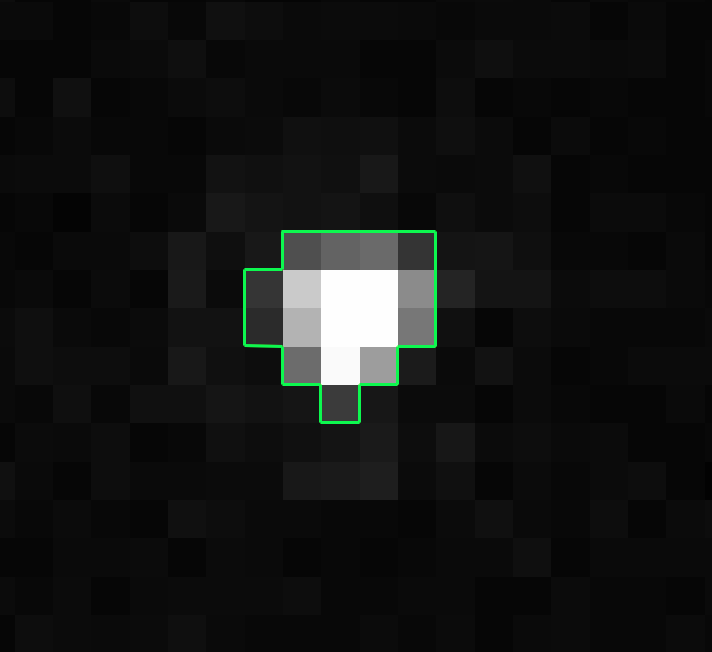}
  \caption{\spitzer\ IRAC2 ($4.5\mu$m) image of K2-3. Pixels inside the green lines are included in the analysis.}
  \label{fig:pixel_selection}
  \bigskip
\end{figure}

Since the positional stability of \spitzer\ has been improved greatly \citep{Grillmair2012}, a relatively few pixels can capture most of the information. Experiments in \cite{Deming2015} typically use a group of 9 pixels contained in a 3$\times$3 pixel square area which covers the stellar image. In practice, we found that the PSF of the images of our sample planets are typically larger than a 3$\times$3 pixel box. To make sure that we include enough pixels to perform PLD correctly, but not too many since each new pixel introduces one more weight parameter into the fitting process, we choose the set of pixels in an observation such that each pixel contributes more than 1\% of the total intensity of the object. An example of our pixel selection is shown in Figure \ref{fig:pixel_selection}. For reproducibility, we list the pixel selection of each AOR of every planet in our sample in the Appendix.

We point out that the PLD method assumes no blending source in selected pixels. A blending primary or secondary star could dilute the transit light curve and result in underestimated planet radius. \cite{Furlan2017} presents the result from the \kepler\ imaging follow-up observation program, and reports that around 10\% of KOI host stars have at least one companion detected within 1\arcsec, while the actual fraction should be even larger due to the limited imaging sensitivities. Furthermore, assuming that planets orbit the primary stars or the brightest companions, they derive an average radius correction factor of 1.06 or 3.09 for KOI planets. Since the \spitzer\ IRAC pixel size is $\sim$1.2\arcsec, in comparison with \kepler's $\sim$4\arcsec pixel size, the blending source problem of \spitzer\ light curves should be considerably smaller than that of \kepler\ light curves. We include a star into our sample only if any catalogued companion/background source within 3\arcsec radius from the primary star is more than 3 magnitudes fainter than the primary star in the $\rm K_s$ band. We estimate that with this criterion, blending sources can only contributes no more than a few percent uncertainty to the transit depth measurements, negligibly small compared to the uncertainty from \spitzer\ photometric noise. As for the blending problem in \kepler\ results, we correct the \kepler\ band effective radius of each planet in our sample with the correction factors presented in \cite{Furlan2017}. 

\subsection{Transit Model and Parameters}\label{sec:SetParameters}


We model the transit light curve using the BAsic Transit Model cAlculatioN (hereafter BATMAN) package provided by \cite{Kreidberg2015}. We fix orbital period, and duration to the values measured by \kepler/\K2, and calculated the semi-major axis using the Kepler's law with the stellar mass and planet orbital periods. To enforce self-consistency, we solve for the orbital inclination using these three parameters rather than employing the reported values. We note that 11 planets in our sample have previously-derived Kepler transit duration too long to match the semi-major axis and period even for 90 degrees inclination (edge-on). This could result from either imprecisely determined transit parameters from previous works, or the fact that the planets have slightly eccentric rather than circular orbits. We believe our sample have low eccentricities as a rule, for reasons we explain below, so we assume an edge-on inclination for these 11 planets and recalculate their orbital semi-major axis according to the literature period and duration values.

Orbital eccentricities are generally poorly determined from transit light curves alone. We elect to set eccentricity to zero for three reasons: first, because tidal theory predicts that M dwarf planets at these orbital distances likely circularized shortly after formation \citep{Barnes2017}. Secondly, half of our planet sample resides in systems with more than one transiting planet, and \cite{Xie2016} measured low orbital eccentricity ($<0.07$) and mutual inclination as a rule for multi-transit systems. Thirdly, ensemble analyses of M dwarf planetary systems show that the singly-transiting planets are dominated (a fraction of 64-84\% according to \cite{Xie2016}) by a dynamically cool population similar to the multi-transit systems \citep{Ballard2016, Moriarty2016}. Employing this reasoning, then of the 28 planets in our sample, it is likely that at least 21 have eccentricities smaller than 0.07 \citep{Xie2016}. 

\cite{Holczer2016} detected significant Transit Timing Variations (hereafter TTVs) in about $1/10$ of all KOIs in their sample, which could produce mid-transit epoch variations as large as a few hundred minutes, and significantly bias our depth fitting results for single transit light curves, therefore we set the transit epoch as another free parameter in our fitting. This enlarges the transit depth uncertainty, but we consider it a source of uncertainty necessary to be taken into account.

To make sure that all planets in our sample are analyzed consistently, we choose \kepler\ band transit depths and transit parameters from uniform literature sources. Two previous works, \cite{Vanderburg2016} and \cite{Crossfield2016}, studied large samples of \K2\ planets, each using an uniform approach. \cite{Vanderburg2016} released their transit analysis results of 234 planet candidates from Campaigns 0--3 of the \K2\ observations. Later \cite{Crossfield2016} analyzed the Campaign 0--4 data of the \K2\ mission, and found 197 planet candidates which were then classified into 3 more detailed categories, and presented their transit parameters. For \K2\ planets in our sample, we use parameters from \cite{Crossfield2016} because it employed a more detailed false positive analysis using supporting high-resolution spectroscopy, and therefore achieved higher transit fitting precision. As a result, for all \K2\ planets in our sample but K2-9b, which could not be validated with high confidence in \cite{Crossfield2016} due to inaccurate stellar parameters, we take planet period, transit duration and \kepler\ band transit depths from \cite{Crossfield2016}'s catalog. Parameters of K2-9b are taken from \cite{Vanderburg2016}.

Similarly, the host star parameters of our sample \K2\ planets are all taken from \cite{Martinez2017}, which studied a large set of \K2\ host stars with high accuracy, except for three stars that are not included in \cite{Martinez2017}'s catalog: K2-26, K2-28, and EPIC205686202. For these three stars, we use stellar parameters from \cite{Crossfield2016}, \cite{Dressing2017} and \cite{Huber2016} respectively.

For planets detected during the original \kepler\ mission, we choose orbital parameters preferentially from \cite{Holczer2016} because it includes detailed TTV and TDV analysis, and all stellar parameters are from \cite{Mann2013}, which includes the most recent accurate analysis of KOI stellar radius and mass.

We model the stellar limb-darkening effect with the quadratic limb-darkening law defined by 2 coefficients. \cite{Claret2011} computed limb-darkening coefficients of different models for a list of photometric systems, and put together a catalogue of coefficients on a hyper grid defined by stellar parameter values and photometric bands. We look up the quadratic limb-darkening coefficients of each star in this catalogue based on its $T_{\rm eff}$, log$g$, and [Fe/H] values from previous works, and set them as input parameters of BATMAN.

\subsection{Gaussian Process}\label{sec:GP}

The PLD method can significantly reduce or eliminate red noise in \spitzer\ photometry \citep{Deming2015}, but to ensure that our transit fitting is not biased by any additional time-correlated noise, we also incorporate a Gaussian process into the fitting. \cite{Luger2016, Luger2017} demonstrated that the PLD+GP technique results in the highest-precision \K2\ photometry. The covariance matrix of our Gaussian process is parameterized with a Matern$-3/2$ kernel, demonstrated to effectively capture residual systematic noise in Spitzer data \citep{Louden2017, Gibson2013, Evans2015}. This kernel is defined by two hyper-parameters: $\rho$, which encodes the autocorrelation timescale, and amplitude $A$, which encodes the fractional contribution from off-diagonal elements to the total covariance. For A = 0, the noise is purely Gaussian.

We test and apply Gaussian Process to our PLD fitting algorithm using the $celerite$ fast GP regression library \citep{celerite}.

\subsubsection{Simulation and Test}

Kepler-138b is a well studied planet in a multi-planetary system with a relatively precisely determined transit depth \citep{Jontof2015,Kipping2014}, and it is hosted by one of the brightest stars in our sample, with a low \spitzer\ photometry noise. We simulate red noise with various correlation time scales using the Matern-3/2 covariance matrix based on the light curve of Kepler-138c, including the transit model and the temporal ramp in detector sensitivity. We then assign the total flux time series into 9 pixels according to the mean and variance of the fractional contribution from each pixel in the real Kepler-138c observation. The assignment ensures that the total flux time series is not changed. 

\begin{figure*}
\begin{center}
    \subfigure
	{%
	\label{fig:first}
	\includegraphics[width=0.45\textwidth]{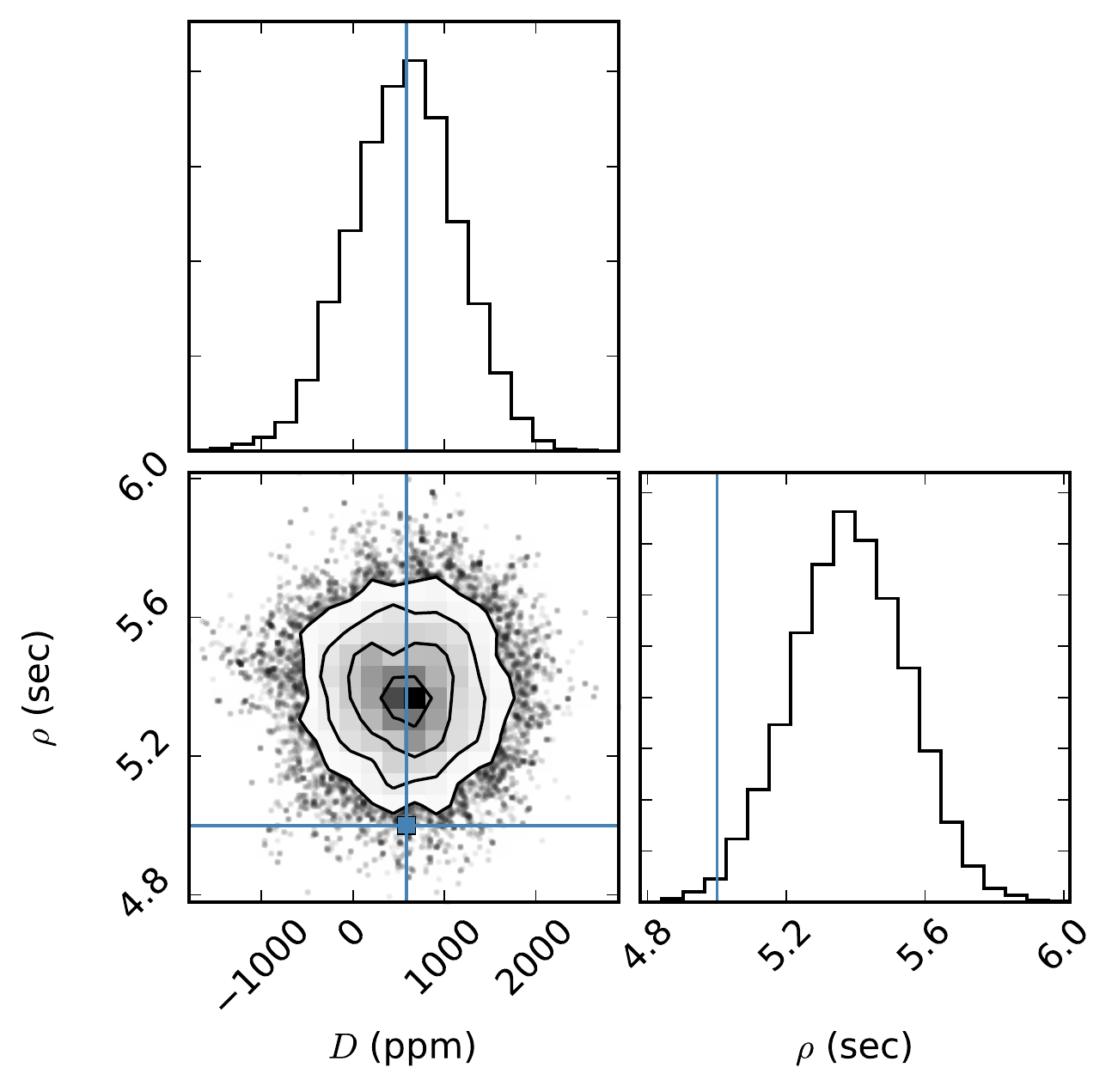}
	}%
    \subfigure
	{%
	\label{fig:first}
	\includegraphics[width=0.45\textwidth]{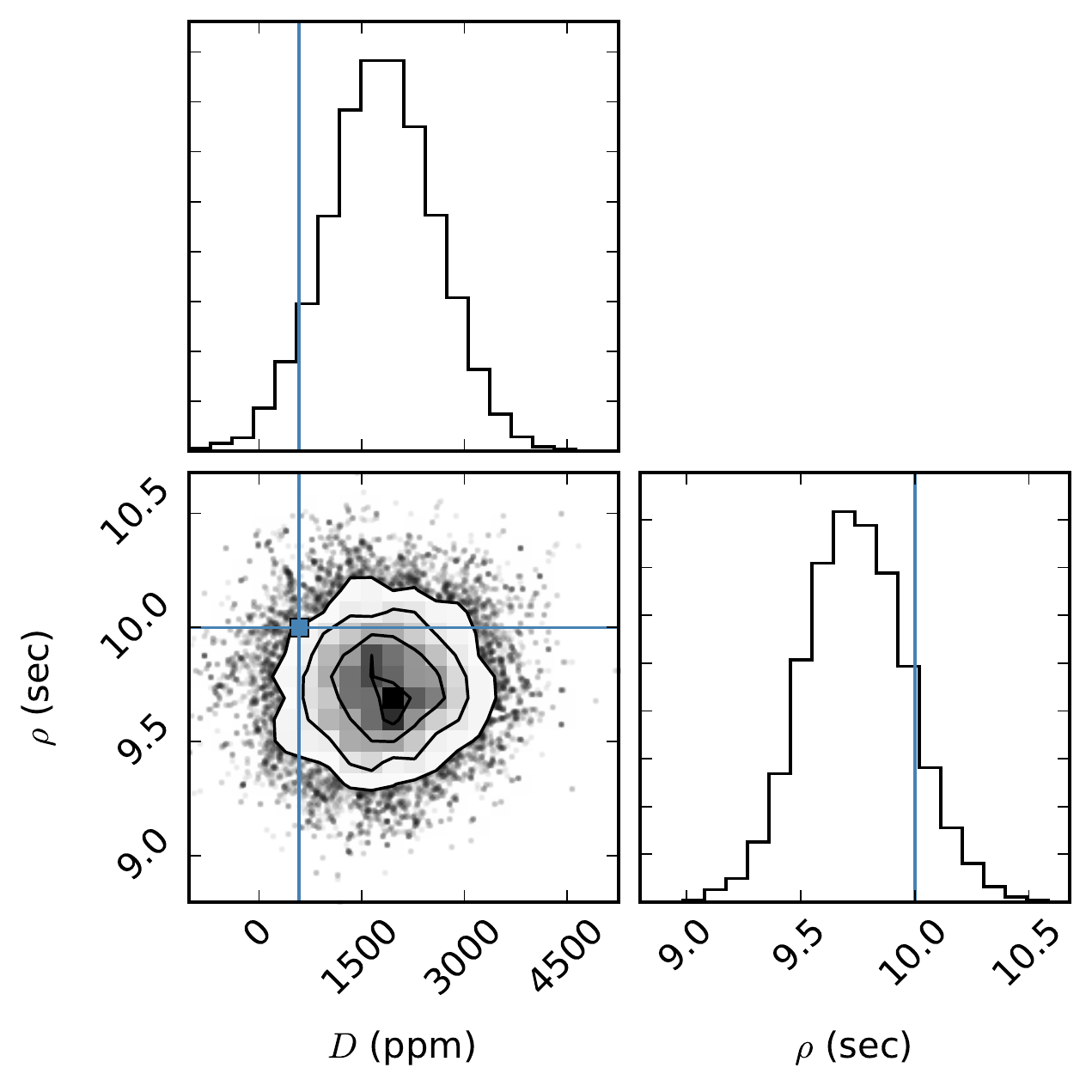}
	}\\
    \subfigure
	{%
	\label{fig:first}
	\includegraphics[width=0.45\textwidth]{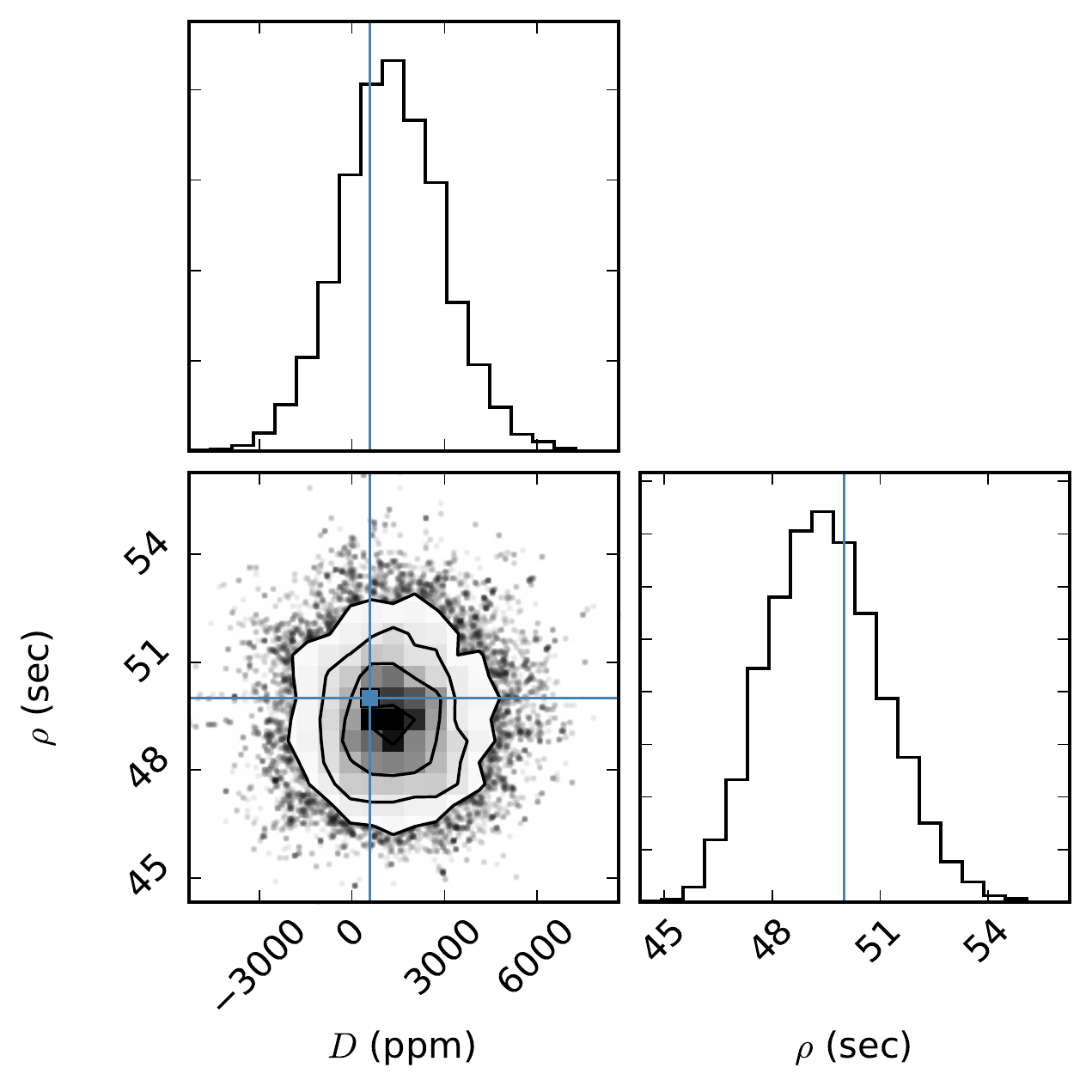}
	}%
    \subfigure
	{%
	\label{fig:first}
	\includegraphics[width=0.45\textwidth]{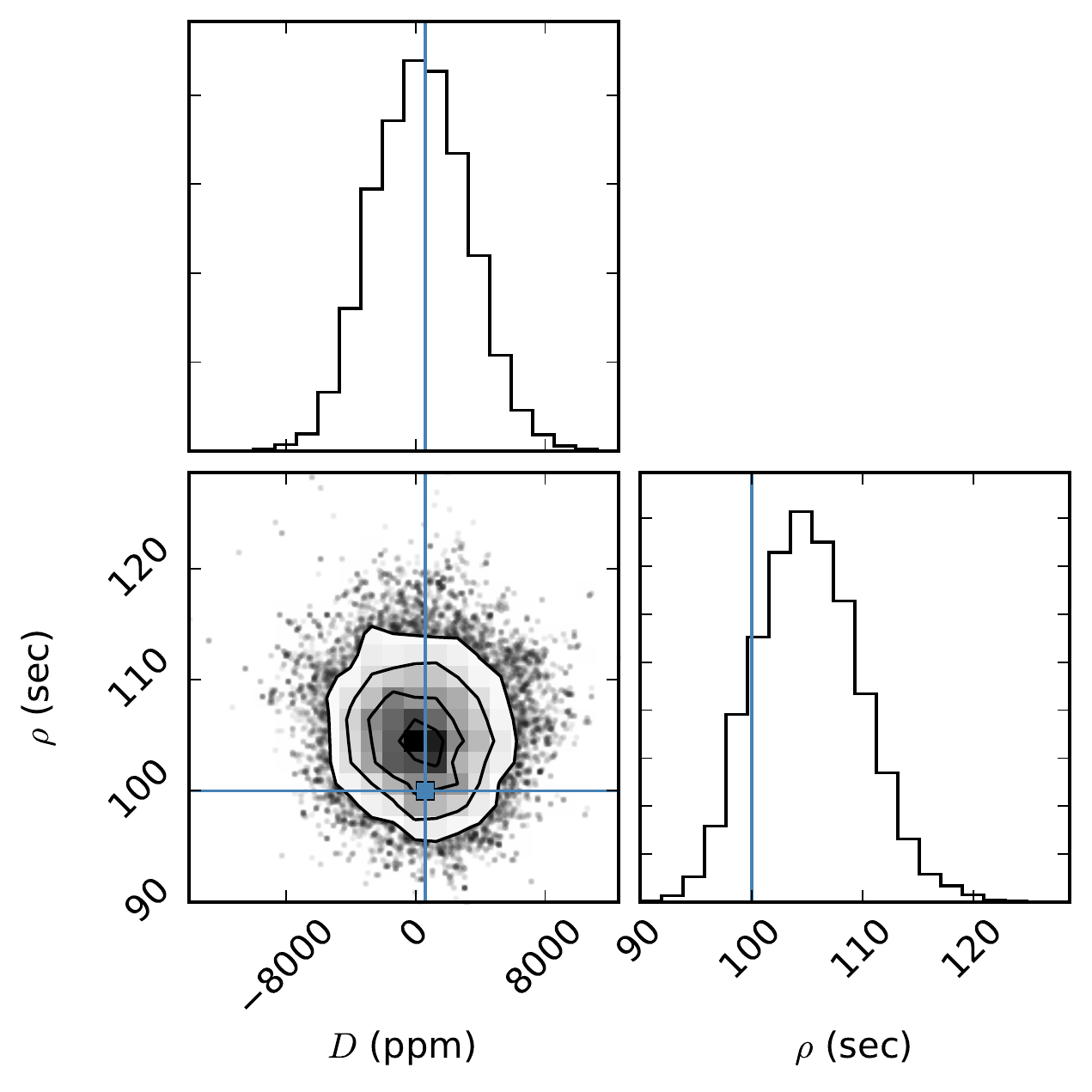}
	}%
\end{center}
\caption{The posterior of the transit depth and the autocorrelation time scale of the Gaussian process tests. The blue vertical lines represent the input transit depth and autocorrelation time scale. The four panels are four cases with different input autocorrelation time scales: 5 seconds on the upper left, 10 seconds on the upper right, 50 seconds on the lower left, and 100 seconds on the lower right. The input transit model has a depth of 586 ppm. We can see that our Gaussian process can recover the input transit depth within $\sim 1 \sigma$ uncertainty. The autocorrelation time scale can mostly be recovered, with the exception of input time scale of 5 seconds, and this is likely due to the fact that the input cadence is $\sim$6 seconds.}
\label{fig:GP_triangle}
\bigskip
\end{figure*}

We then apply our PLD and GP combined fitting pipeline to the simulated Kepler-138c time series, and compare the recovered correlation time scales and transit depths with the input values. We test 4 different correlation time scales: 5 seconds, 10 seconds, 50 seconds, and 100 seconds, and the results are shown in Figure \ref{fig:GP_triangle}. We can see from Figure \ref{fig:GP_triangle} that the recovered time scales are in general consistent with the input values. In the case when the input time scale is 5 seconds, the posterior converges to a value slightly larger than 5 seconds. This is because the light curve cadence is around 6 seconds, and this feature biases the GP to a time scale between 5 and 6 seconds. More importantly, the posteriors of transit depth are all symmetrically distributed around the input value, and consistent within 1 $\sigma$ uncertainty. This proves that our GP can successfully recover a transit signal similar to Kepler-138c when a combination of red and white noise are present. We also notice a positive correlation in our test results between the uncertainty on transit depth and the red noise time scale: The longer the correlation length, the greater the measurement uncertainty. 


\subsubsection{Apply GP to the Sample}

We combine the PLD and GP into our MCMC fitting pipeline, and apply the pipeline to each observation of every planet in our sample. The results show that all \spitzer\ light curves in our sample are dominated by white noise: the noise correlation time scales are all smaller than the cadence of the observations. From our previous test, we know that the best-fit GP kernel tends to be shorter than the observational cadence if the contribution of red noise is low. We conclude, similarly to \cite{Deming2015}, that PLD adequately removes instrumental noise without introducing systematic error. The fitting results of other transit parameters are presented in the next section.

\bigskip

\section{Results}\label{sec:results}

We summarize our fitting result of the transit depth of each planet in Table \ref{tab:result}. Also listed are radii of host stars and transit depths in the \kepler\ band for future references. In addition, we present the mid-transit time of all 64 transits in the Appendix, which will allow more efficient scheduling of \HST\ and \JWST\ observations to further characterize our sample in the future.

\begin{figure}
 \includegraphics[width=\linewidth]{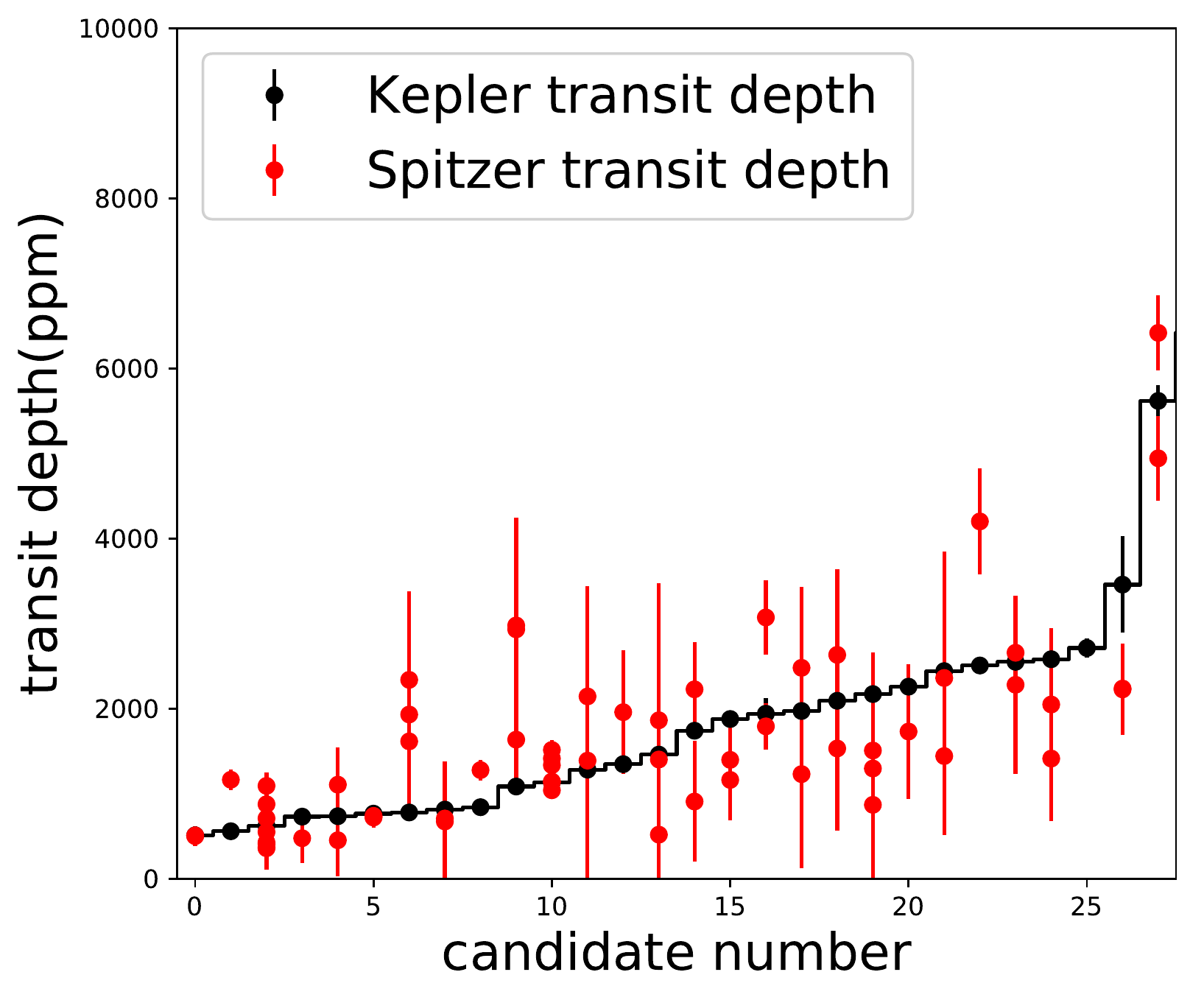}
  \caption{Transit depths of the 28 planets that we analyzed, ordered according to their \kepler\ band transit depths, which are plotted as black points with error bars. In comparison, the \spitzer\ band transit depth of every transit of each planet with uncertainty is over plotted in red. For most planets, their \kepler\ and \spitzer\ transit depths are consistent within $1\sigma$, with a few exceptions that will be discussed in section \ref{sec:atmosphere}.}
  \label{fig:compare_aor}
  \bigskip
\end{figure}

Figure \ref{fig:compare_aor} shows a comparison of the \kepler\ and \spitzer\ transit depths for each planet, in increasing order with \kepler\ transit depth. In black are the \kepler\ transit depths (with individual errors too small to be seen in this scale), where red error bars show the \spitzer\ transit depths for each observation. We find that multiple \spitzer\ transit observations of the same planet produce consistent results within 1 or 2 sigma for each planet, an independent piece of evidence that we are achieving close to Poisson-limited precision on the transit depth. 

\begin{figure}
 \includegraphics[width=\linewidth]{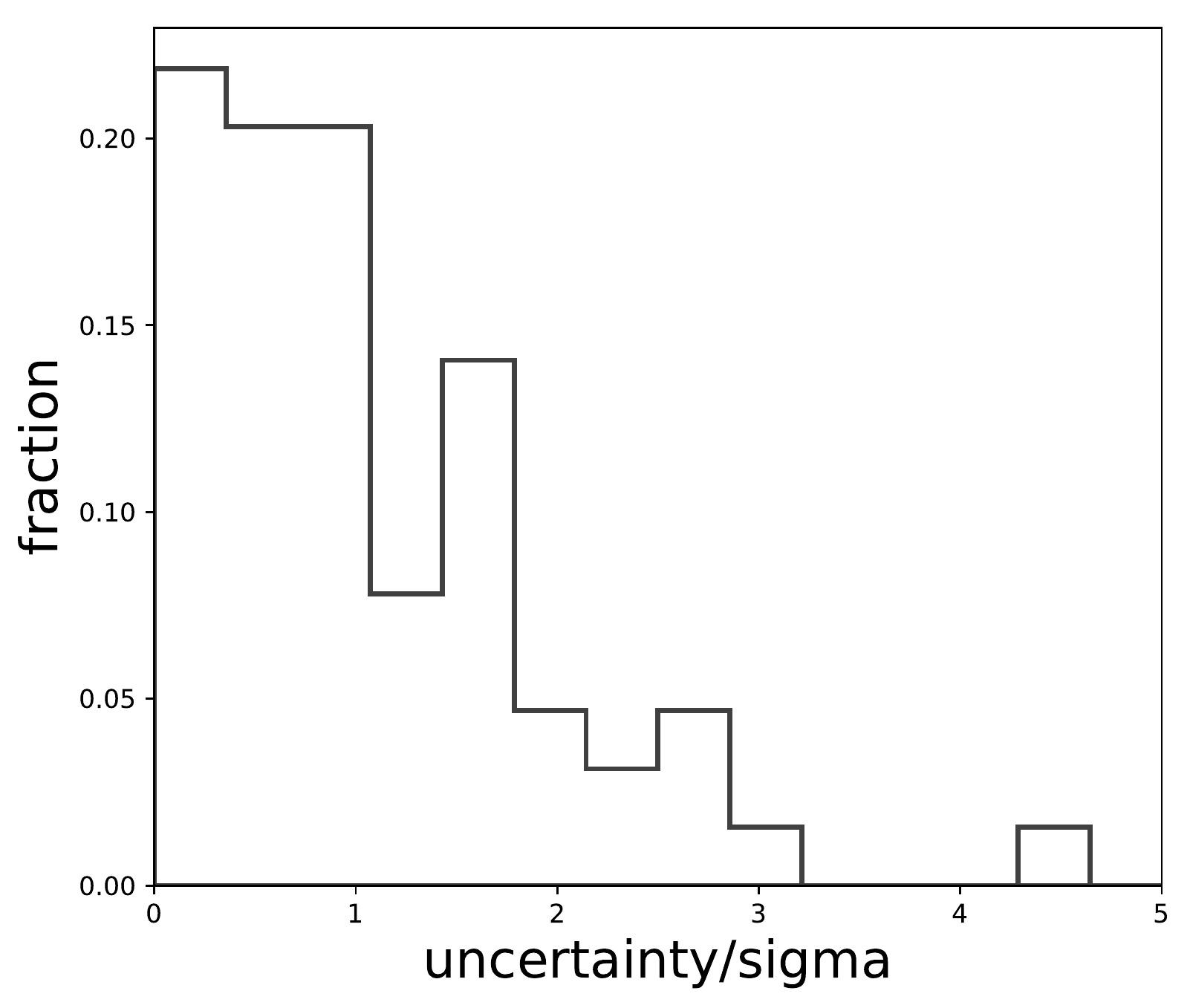}
  \caption{The distribution of discrepancies between \spitzer\ band and \kepler\ band transit depths of the 28 planets that we analyzed, scaled with their transit depth uncertainties. Around 61\% of the transit measurements have discrepancies within 1$\sigma$, and around 98\% observations have discrepancies within 3$\sigma$.}
  \label{fig:sigma_aor}
  \bigskip
\end{figure}

On Figure 5 we show the fractional distribution of the discrepancy between Kepler transit depths and Spitzer transit depths, scaled to the number of sigma between them. Around 61\% observations have discrepancies within 1 sigma, and about 98\% observations have discrepancies within 3 sigma. 

\begin{figure}
 \includegraphics[width=\linewidth]{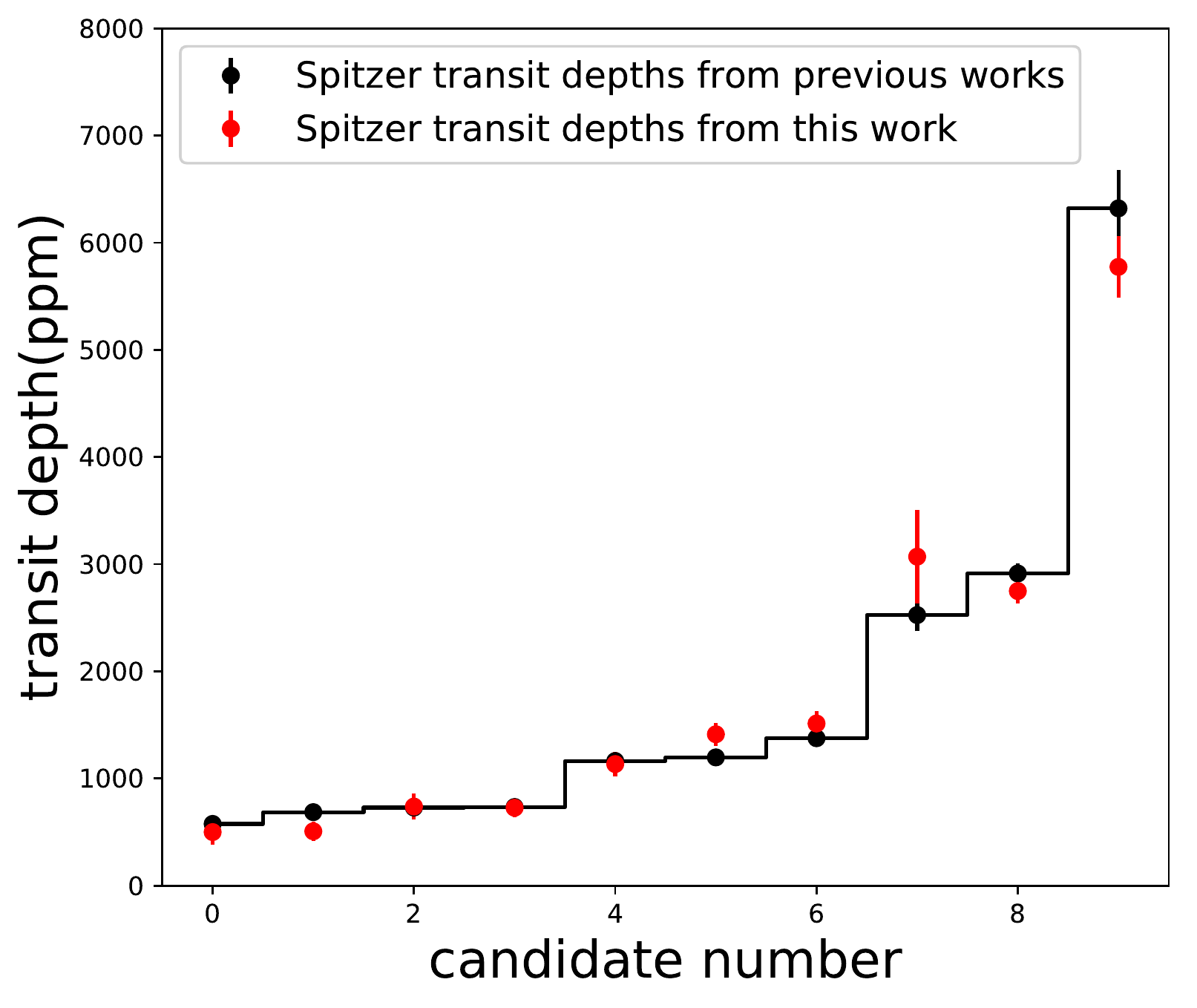}
  \caption{A comparison between the depths of 10 transits measured in this works and previous works, ordered according to the transit depths reported in previous works. The differences in all 10 transits are smaller or around 1~$\sigma$. All data points have error bars, but for several of them, the error bar sizes are smaller then the marker sizes.}
  \label{fig:compare_previous}
  \bigskip
\end{figure}

Previous works have measured the transit depths on the \spitzer\ 4.5~$\mu$m band for six planets in our sample using the PLD method. \cite{Beichman2016} analyzed 3 transits of K2-3b, 2 transits of K2-3c, 2 transits of K2-3d and 1 transit of K2-26b, 
\cite{Benneke2017} and \cite{Chen2018} studied 1 transit of K2-18b and K2-28b respectively. In this work, we reanalyzed all transits measured in previous works, and compared our result with the previous result of each transit (10 transits in total) in Figure \ref{fig:compare_previous}. The differences in all transit depths are smaller or around 1~$\sigma$, and the uncertainties on our measurements are larger because we modeled the photometric noise with a Gaussian process, and included 2 extra parameters describing the kernel used in the Gaussian process. Two reasons could lead to the differences between our results and previous results: first, differences in the selections of input stellar parameters, and second, differences in the \spitzer\ PSF selection. To ensure sample consistency and facilitate valid comparison with the \kepler\ band transit depths, we have used revised planet parameters from the \K2\ analysis in \cite{Crossfield2016} and the stellar parameters from \cite{Martinez2017} for the 3 planets with differences larger than 1~$\sigma$. And we have included all pixels around the star that contribute more than 1\% of the total flux to ensure that our PLD analysis is correct, while \cite{Beichman2016} and \cite{Chen2018} chose circular PSF of 2.2 and 1.9 pixels in radius respectively, and \cite{Benneke2017} used a set of 3$\times$3 pixels. The pixel selections of all targets in our sample are listed in the appendix.

Combining with all available new transit data that were not analyzed before, we updated the transit depths on the \spitzer\ 4.5~$\mu$m channel for the six planets, and listed the results from this work and previous works in Table \ref{tab:result_compare} for future references. 

\begin{table*}
\renewcommand*{\arraystretch}{1.7}
    \caption{Summary on Fitting Results of 28 Planets}
    \centering
    \begin{tabularx}{\textwidth}{nnnnnnnnn}
    \hline\hline
    Name & $R_{\rm s}$\footnote{The radius with uncertainty of the host star.} ($R_{\sun}$) & $D_{Kep}$\footnote{The \kepler\ transit depth from previous works. Details can be found in section \ref{sec:SetParameters}.} (\rm ppm) & $\sigma_{Kep}$\footnote{The uncertainty on the \kepler\ transit depth from previous works.} (\rm ppm) & $D_{Spitzer}$ (\rm ppm) & $\sigma_{Spitzer}$ (\rm ppm) & $\frac{\Delta R_{\rm p}}{R_{\rm p}}$ & IRAC Channel \\
    \hline
    KOI247.01 & 0.547$\pm$0.034 & 1083 & 30 & 2810 & 863 & 0.15 & 4.5 $\mu$m \\
    Kepler-49b & 0.583$\pm$0.034 & 1971 & 30 & 1758 & 754 & 0.21 & 4.5 $\mu$m \\
    Kepler-49c & 0.583$\pm$0.034 & 1458 & 30 & 1410 & 890 & 0.32 & 4.5 $\mu$m \\
    Kepler-504b & 0.413$\pm$0.039 & 1877 & 30 & 1300 & 391 & 0.15 & 4.5 $\mu$m \\
    Kepler-26c & 0.604$\pm$0.034 & 2170 & 30 & 1118 & 998 & 0.45 & 4.5 $\mu$m \\
    Kepler-125b & 0.517$\pm$0.035 & 2580 & 30 & 1790 & 612 & 0.17 & 4.5 $\mu$m \\
    KOI252.01 & 0.549$\pm$0.034 & 2440 & 30 & 2101 & 1102 & 0.26 & 4.5 $\mu$m \\
    KOI253.01 & 0.562$\pm$0.034 & 2257 & 30 & 1730 & 791 & 0.23 & 4.5 $\mu$m \\
    Kepler-505b & 0.562$\pm$0.034 & 2506 & 30 & 4200 & 620 & 0.07 & 4.5 $\mu$m \\
    Kepler-138c & 0.553$\pm$0.034 & 729 & 30 & 475 & 291 & 0.31 & 4.5 $\mu$m \\
    Kepler-138d & 0.553$\pm$0.034 & 621 & 30 & 648 & 96 & 0.07 & 4.5 $\mu$m \\
    Kepler-205c & 0.622$\pm$0.035 & 734 & 30 & 785 & 305 & 0.19 & 4.5 $\mu$m \\
    Kepler-236c & 0.551$\pm$0.034 & 1281 & 30 & 1723 & 991 & 0.29 & 4.5 $\mu$m \\
    Kepler-249d & 0.456$\pm$0.037 & 812 & 30 & 687 & 481 & 0.35 & 4.5 $\mu$m \\
    Kepler-737b & 0.525$\pm$0.035 & 1739 & 30 & 1408 & 488 & 0.17 & 4.5 $\mu$m \\
    Kepler-32d & 0.529$\pm$0.035 & 2091 & 30 & 2104 & 698 & 0.17 & 4.5 $\mu$m \\
    Kepler-36b & 0.630$\pm$0.035 & 1346 & 30 & 1957 & 725 & 0.19 & 4.5 $\mu$m \\
    Kepler-395c & 0.567$\pm$0.034 & 778 & 30 & 2002 & 572 & 0.14 & 4.5 $\mu$m \\
    K2-3b & 0.565$\pm$0.061 & 1134 & 97 & 1306 & 47 & 0.02 & 4.5 $\mu$m \\
    K2-3c & 0.565$\pm$0.061 & 764 & 85 & 728 & 70 & 0.05 & 4.5 $\mu$m \\
    K2-3d & 0.565$\pm$0.061 & 509 & 65 & 503 & 80 & 0.08 & 4.5 $\mu$m \\
    K2-9b & 0.366$\pm$0.053 & 3457 & 567 & 2229 & 373 & 0.08 & 4.5 $\mu$m \\
    K2-18b & 0.411$\pm$0.053 & 2549 & 200 & 2749 & 118 & 0.02 & 4.5 $\mu$m \\
    K2-21b & 0.721$\pm$0.059 & 557 & 63 & 1163 & 122 & 0.05 & 4.5 $\mu$m \\
    K2-21c & 0.721$\pm$0.059 & 840 & 100 & 1276 & 121 & 0.05 & 4.5 $\mu$m \\
    K2-26b & 0.51$\pm$30 & 1939 & 186 & 2716 & 326 & 0.06 & 4.5 $\mu$m \\
    K2-28b & 0.280$\pm$0.031 & 6417 & 200 & 5774 & 287 & 0.02 & 4.5 $\mu$m \\
    EPIC 210558622 & 0.673$\pm$0.038 & 5617 & 187 & 5584 & 341 & 0.03 & 4.5 $\mu$m \\
    \hline
    Kepler-138d & 0.553$\pm$0.034 & 621 & 30 & 457 & 137 & 0.15 & 3.6 $\mu$m \\
    K2-18b & 0.411$\pm$0.053 & 2549 & 200 & 2656 & 107 & 0.02 & 3.6 $\mu$m \\
    \hline
    \end{tabularx}
    \label{tab:result}
\bigskip
\end{table*}

\begin{table*}
\renewcommand*{\arraystretch}{1.7}
    \caption{Compare with Previous \spitzer\ Transit Depth Measurements}
    \centering
    \begin{tabularx}{\textwidth}{nnnnnnc}
    \hline\hline
    Name & $D_{\rm previous~works}$\footnote{Transit depths on the \spitzer\ band from previous works.} (ppm) & $N_{\rm previous~work}$\footnote{Number of transits used in previous works.} & $D_{\rm this~work}$\footnote{Transit depths on the \spitzer\ band measured in this work.} (ppm) & $N_{\rm this~work}$\footnote{Number of transits used in this work.} & Difference\footnote{Differences between the transit depths from this work and prevous works.} ($\sigma$) & Literature \\
    \hline
    K2-3b & $1246\pm 78$ & 3 & $1306 \pm 47$ & 6 & 0.66 & \cite{Beichman2016} \\
    K2-3c & $732\pm 27$ & 2 & $728 \pm 70$ & 3 & 0.07 & \cite{Beichman2016} \\
    K2-3d & $623\pm 25$ & 2 & $503\pm 80$ & 2 & 1.43 & \cite{Beichman2016} \\
    K2-18b & $2913\pm 94$ & 1 & $2749\pm 118$ & 1 & 1.09 & \cite{Benneke2017} \\
    K2-26b & $2525\pm 150$ & 1 & $3070\pm 438$ & 2 & 1.18 & \cite{Beichman2016} \\
    K2-28b & $6320\pm 358$ & 1 & $5774\pm 287$ & 1 & 1.19 & \cite{Chen2018} \\
    \hline
    \end{tabularx}
    \label{tab:result_compare}
\bigskip
\end{table*}


\bigskip

\section{Atmosphere properties}\label{sec:atmosphere}

With transit depths measured only in a single optical and infrared bandpass for an exoplanet, we have only a hint of the atmospheric properties. We cannot probe the molecular content of the atmosphere, like studies with higher-resolution transmission spectroscopy. However, a difference in transit depth with wavelength does necessitate the existence of an atmosphere of some kind. This alone is an uncertain proposition for planets at this radius and temperature range orbiting M dwarfs, for which tidal locking and strong stellar magnetic activity pose danger to atmospheres \citep{Zendejas2010, Heng2012, Barnes2013}. Additionally, we already know from \cite{Sing2016} for Jovian planets that the difference in transit depth between the blue-optical and mid-infrared is correlated with the presence of water absorption. 

For the purpose of this study, we quantify this difference between the optical and infrared transit depth with a power law slope $\alpha$ between the two. 

Assuming a hydrogen dominated atmosphere in hydrodynamic equilibrium, \cite{Lecavelier2008} derived the variation of the planet effective radius as a function of the observation wavelength on transmission spectra: $\alpha H = dR_{\rm p}(\lambda)/d\ln{\lambda}$ (which will be described later as equation \ref{R_effect}), where the critical parameter $\alpha$ is the index of the power law describing the scattering cross section distribution of atmosphere particles as a function of wavelength. According to light scattering theories, the value of $\alpha$ is a useful indicator of particle sizes should a haze/cloud feature is present, and the efficiency of scattering decreases as photons approach the size of scattering particles \citep{Hansen1974}. In particular, we are familiar with the Rayleigh scattering, which yields $\alpha=-4$ and is the key mechanism governing the scattering of small particles which produces haze features on transmission spectra. Another special case is the flat transmission spectra with molecular features obscured. These are caused by high altitude cloud decks on planet atmospheres, and the clouds consist of large particles by which stellar light is scattered through a Mie scattering process which has $\alpha=0$. Therefore it is in theory possible to constrain the particle size in planet atmospheres with this cross section power law index $\alpha$, and $\alpha$ could be measured from the \kepler\ and \spitzer\ band transit depths.

\subsection{Haze/Cloud Power Law Slope}

\cite{Fortney2005} showed that the optical depth $\tau$ for lights of wavelength $\lambda$ through a planet's atmosphere at an altitude $z$ in the line of sight can be described by the formula 

\begin{equation}\label{optical_depth}
\tau (\lambda, z) \approx \sigma (\lambda)n(z)\sqrt{2\pi R_p H},  
\end{equation}

\noindent
where $n(z)$ is the number density of particles in the atmosphere with absorption cross section $\sigma (\lambda)$, $R_{\rm p}$ is the planet radius, and $H$ is the local atmosphere scale height which describes the vertical profile of particle density: $n(z) = n_{(z=0)}exp(-z/H)$. The absorption cross section varies according to a power law in wavelength: 

\begin{equation}\label{scattering}
\sigma (\lambda) \sim \lambda^{\alpha}, 
\end{equation}

\noindent
where the value of $\alpha$ depends on the size of atmosphere particles. For Rayleigh scattering which involves small particles, $\alpha = -4$, and for Mie scattering which is caused by large particles, $\alpha = 0$. 

The effective planet radius $R(\lambda)$ for the star light at wavelength $\lambda$ is defined such that a sharp occulting disk of this radius would produce the same transit depth as the real planet with with translucent atmosphere \citep{Lecavelier2008}. Therefore, we can integrate the amount of light on wavelength $\lambda$ absorbed over the entire atmosphere using equation \ref{optical_depth}, and evaluate the effective radius $R(\lambda)$. Using equation \ref{optical_depth} and \ref{scattering}, we can get the equation describing the effective radius as a function of wavelength:

\begin{equation}\label{R_effect}
\alpha H = \frac{dR_{\rm p}(\lambda)}{d\ln{\lambda}}.
\end{equation}

\begin{figure}
 \includegraphics[width=\linewidth]{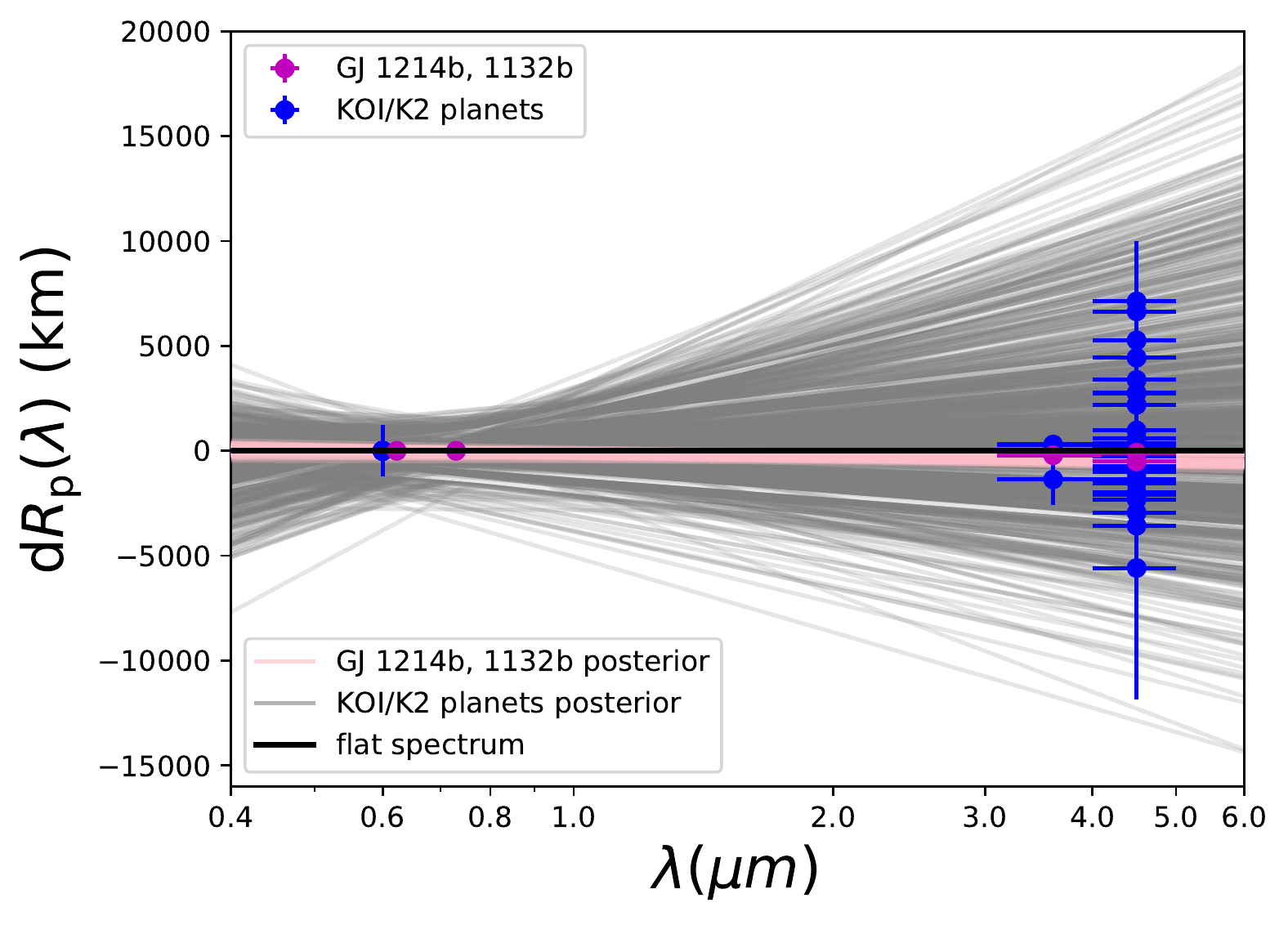}
  \caption{Posteriors of the linear fits for all 30 planets in our sample based on their \kepler\ band and combined \spitzer\ band transit depths. The blue data points are the 28 KOI/K2 planets, and the magenta data points are the two bright previously studied planets: GJ 1214b and GJ 1132b. The grey lines are the posteriors of the linear fits of KOI/K2 planets, and the pink lines are the posteriors of the linear fits of GJ 1214b and GJ 1132b. The black line represents a flat transmission spectrum.}
  \label{fig:transmission_km}
  \bigskip
\end{figure}

\noindent
We fit the large scale transmission spectrum slope $dR_{\rm p}(\lambda)/d\ln{\lambda}$ of each planet in our sample using the two transit depth measurements of each planet on the \kepler\ band and the \spitzer\ band from section \ref{sec:results}, and the transit depth uncertainties are taken into account during the linear fitting. The center values and uncertainties of $dR_{\rm p}(\lambda)/d\ln{\lambda}$ are listed in Table \ref{tab:slope}. The fitting posteriors of each planet are shown in Figure \ref{fig:transmission_km}, and we added the data and corresponding fitting posteriors of two well known super-Earths, GJ 1214b and GJ 1132b, to complement our sample. 
All data and posterior lines are shifted on the figure so that the relative altitude on the \kepler\ band is zero for easy comparison. We can see that the posterior lines are partly distributed around a flat spectrum, but there is also a small portion of transmission spectra slopes deviated from the flat spectrum, and featuring positive values. We will discuss more about the bimodal distribution in the following sections. 

With our measurements of $dR_{\rm p}(\lambda)/d\ln{\lambda}$ in hand, we have a constraint on the product $\alpha H$. We now require an estimate of scale height to extract a constraint on $\alpha$. 

\subsection{Atmosphere Scale Height}

Assuming an ideal gas in hydrostatic equilibrium, the pressure of the atmosphere $P$ as a function of height $z$ is given by $d \ln p = -1/H dz$. Here $H$ is defined to be the “scale height.” At this height, the atmospheric pressure has fallen from the base pressure of $P_0$ to $P_0e^{-1}$. H is expressed as follows: 

\begin{equation}\label{scale_height}
H = \frac{kT}{\mu g},
\end{equation}

\noindent
where $k$ is the Boltzmann's constant, and $T$, $\mu$ and $g$ are the local temperature, mean molecular weight and gravity respectively. For a hydrogen dominated atmosphere, $\mu=2$, while often $\mu=2.3$ is used in transmission spectra analysis assuming H/He dominated atmosphere \citep{DeWit2013, Lecavelier2008}. Since we only have transit observations on two wavelength bands, and the \spitzer\ band transit uncertainties are large, our data quality doesn't allow high mean molecular weight atmosphere analysis. Thus we adopt $\mu=2$, which could enlarge the atmosphere scale height and lead to smaller index $\alpha$ values. Because we apply uniform $\mu$ values to all planets in our sample, which also all have similar sizes and occupy similar positions in their systems, the statistical distribution of $\alpha$ values will still reflect the ensemble property of the population. 

The surface gravity of a planet with mass $M_p$ and radius $R_p$ is $g=GM_p/R_p^2$. However, for most of our sample, no dynamical mass estimate exists. We use the mass-radius relation for sub-Neptune sized planets from \cite{Wolfgang2016}: $M/M_{\oplus} = 2.7(R/R_{\oplus})^{1.3}$, to determine planet masses. And for the temperature $T$ in equation \ref{scale_height}, we take the most recent planet equilibrium temperatures from the \it{NASA Exoplanet Archive}. \rm With these parameters assigned, we are able to calculate the approximate atmosphere scale height of each planet in our sample. The scale heights are listed in Table \ref{tab:slope} for future reference.

\subsection{Uncertainty on the Power Law Index $\alpha$}

Using equation \ref{R_effect}, and combining the measurements on $H$ and $dR_{\rm p}(\lambda)/d\ln{\lambda}$, we arrive at the haze/cloud scattering power law index $\alpha$, as presented in Table \ref{tab:slope}, along with the uncertainties on $\alpha$. We propagate our uncertainty on the scale height into our uncertainty on $\alpha$ in the following way. We assume normal distributions for $T_{\rm p}$ and $R_{\rm p}$ with their standard deviations taken from the \it NASA Exoplanet Archive\rm, which are also listed in Table \ref{tab:slope}.

For the planet mass $M_{\rm p}$, we used the result from \cite{Wolfgang2016}, which presented a probabilistic $M_{\rm p}-R_{\rm p}$ relation for sub-Neptune-sized planets ($R_{\rm p}<4R_{\oplus}$): 

\begin{equation}\label{MRrelation}
M/M_{\oplus} = 2.7(R/R_{\oplus})^{1.3}.
\end{equation}

\noindent
The relation was evaluated using a number of previous works on planet mass and radius measurements. However, this relation has a relatively large scatter in mass of $\sim 1.9 M_{\oplus}$, which equals to around 29\% of the mass for a $2R_{\oplus}$ sized planet, and around 42\% of the mass for a $1.5R_{\oplus}$ sized planet. This uncertainty is also propagated into the uncertainty on $\alpha$, and more discussions about planet mass uncertainties can be found in section \ref{sec:mass_discussion}.

\subsection{Power Law Index versus Planet Physical Properties}\label{sec:alpha_property}

\begin{figure}
 \includegraphics[width=\linewidth]{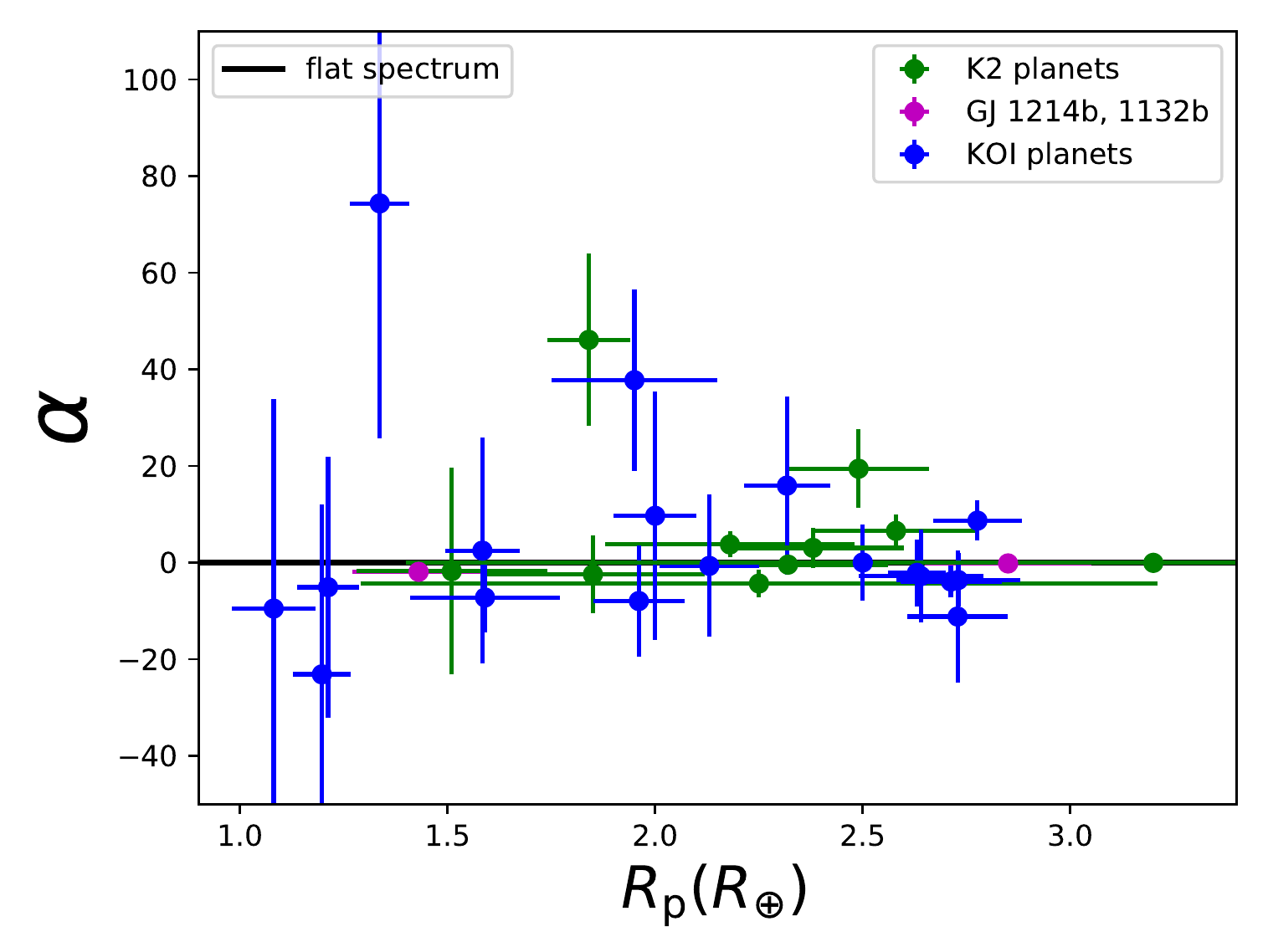}
  \caption{Index $\alpha$ as a function of $R_{\rm p}$ for all 30 planets in our sample. The black line represents a flat transmission spectrum. There is a hint of two populations of planets: one group with lower $\alpha$ values and are distributed around the flat spectrum, the other group with higher $\alpha$ values which consists of only a few planets. And the planets with large $\alpha$ seem to have smaller radius, which is not clear due to the small number of planets.}
  \label{fig:alpha_Rp}
  \bigskip
\end{figure}

\begin{figure}
 \includegraphics[width=\linewidth]{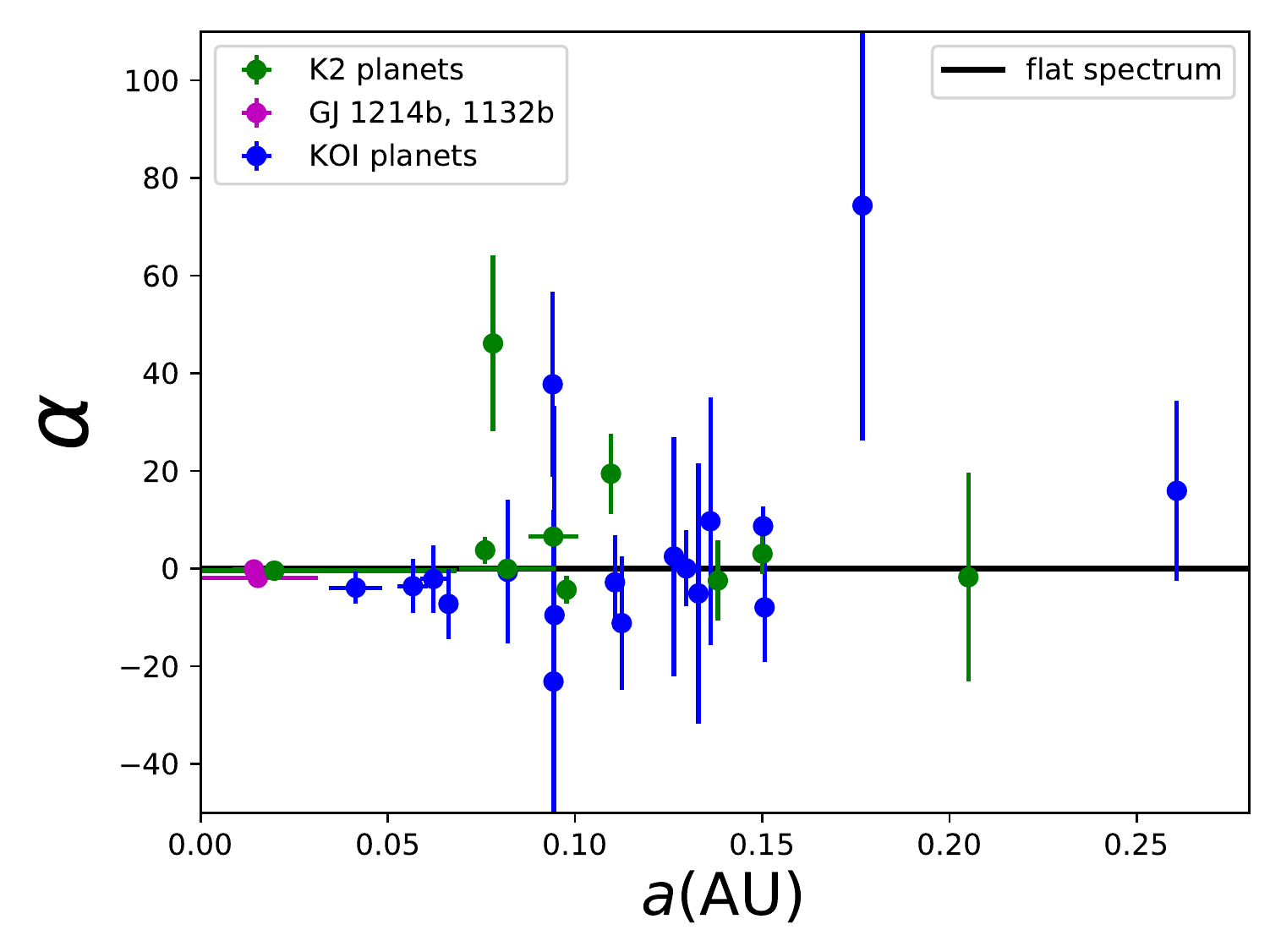}
  \caption{Index $\alpha$ as a function of semi-major axis $a$. The notations are the same as Figure \ref{fig:alpha_Rp}. A hint of two populations of planets can still be observed, But no obvious trend between $\alpha$ values and semi-major axis is found.}
  \label{fig:alpha_SMA}
  \bigskip
\end{figure}

\begin{figure}
 \includegraphics[width=\linewidth]{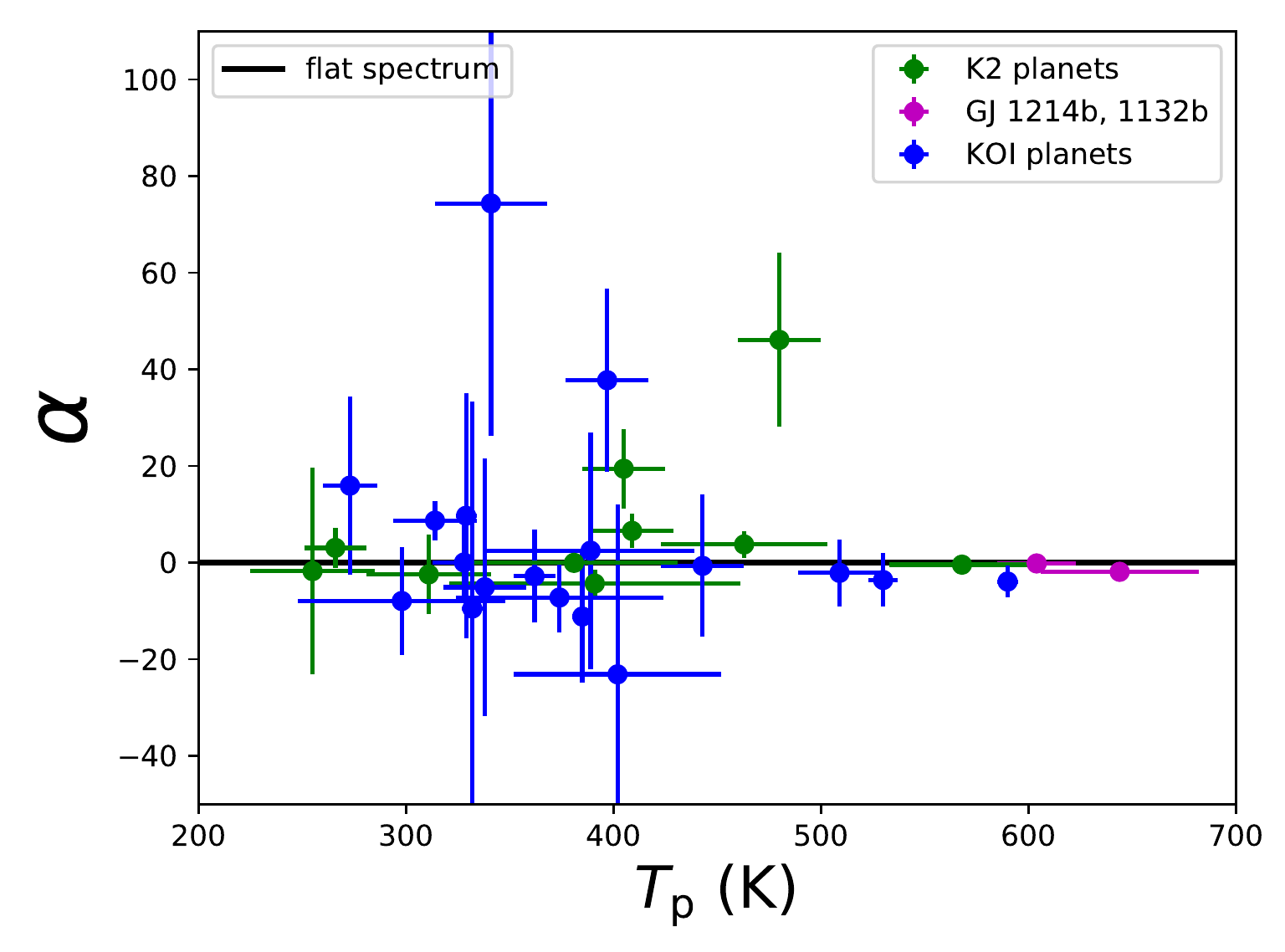}
  \caption{Index $\alpha$ as a function of planet equilibrium temperature $T_{\rm p}$. The notations are the same as Figure \ref{fig:alpha_Rp}. Comparing to the hint of a trend seen in Figure \ref{fig:alpha_SMA}, it is more clear here that all planets with high $\alpha$ values are on the cooler part of the sample--with $T_{\rm p}<500~K$. Discussion about the trend seen in this Figure and Figure \ref{fig:alpha_Rp} and \ref{fig:alpha_SMA} can be found in section \ref{sec:alpha_property} and \ref{sec:2population}.}
  \label{fig:alpha_Tp}
  \bigskip
\end{figure}

We plot $\alpha$ against $R_{\rm p}$, semi-major axis $a$, and equilibrium temperature $T_{\rm p}$, as are shown in Figure \ref{fig:alpha_Rp}, \ref{fig:alpha_SMA} and \ref{fig:alpha_Tp} respectively. The black flat lines represent a flat transmission spectrum, for which $\alpha = 0$. 


We report three main findings. First, the measured values for $\alpha$ mostly cluster around a value indistinguishable from zero. However, there exist a number of planets for which the slope is inconsistent with zero and positive. In the next section, we discuss the statistical likelihood that all measurements are drawn from the same parent population. Secondly, this distribution is traced by the \kepler\ planets and by the \K2\ planets: a population of $\alpha$ values near zero, with a subset of positive values. And thirdly, no physical parameter obviously traces $\alpha$.  

On closer look, Figure \ref{fig:alpha_Rp} indicates that higher $\alpha$ values tend to appear for planets with slightly smaller radius, though there is no clear overall trend. In Figure \ref{fig:alpha_SMA}, we can see that planets orbiting with larger semi-major axis have more scattered $\alpha$ values with larger uncertainty, but no obvious trend is observed. In comparison, a potential correlation shows up when we plot $\alpha$ against planet equilibrium temperatures: Figure \ref{fig:alpha_Tp} shows that large $\alpha$ values appear only for planets around or cooler than 500~K in our sample. The semi-major axis of planets are calculated using their periods and host stars' mass, and equilibrium temperatures are obtained from previous works. 

\begin{table*}
\renewcommand*{\arraystretch}{1.7}
    \caption{Physical, Orbital and Atmospheric Properties of Planets}
    \centering
    \begin{tabularx}{\textwidth}{cccnnncccnnn}
    \hline\hline
    Name & $R_{\rm p}$ & $\sigma_{R_{\rm p}}$ & $a$ & $\sigma_{a}$ & $T_{\rm p}$ & $\sigma_{T_{\rm p}}$ & $dR_{\rm p}(\lambda)/d\ln{\lambda}$ & $\sigma_{dR_{\rm p}(\lambda)/d\ln{\lambda}}$ & $H$ & $\alpha$ & $\sigma_{\alpha}$ \\
     & ($R_{\oplus}$) & ($R_{\oplus}$) & (AU) & (AU) & (K) & (K) & (km) & (km) & (km) &  &  \\
    \hline
    KOI247.01 & 1.95 & 0.20 & 0.094 & 0.001 & 397 & 20 & 3877 & 1470 & 105.0 & 37.73 & 18.85 \\
    Kepler-49b & 2.63 & 0.07 & 0.062 & 0.003 & 509 & 20 & -296 & 1032 & 153.5 & -2.15 & 6.93 \\
    Kepler-49c & 2.13 & 0.12 & 0.082 & 0.002 & 443 & 20 & 63 & 1667 & 153.8 & -0.69 & 15.02 \\
    Kepler-504b & 1.59 & 0.18 & 0.066 & 0.003 & 374 & 50 & -659 & 468 & 106.4 & -7.24 & 7.25 \\
    Kepler-26c & 2.72 & 0.12 & 0.112 & 0.001 & 385 & 5 & -1470 & 1517 & 148.4 & -11.19 & 13.65 \\
    Kepler-125b & 2.67 & 0.12 & 0.041 & 0.007 & 590 & 5 & -795 & 574 & 177.0 & -3.94 & 3.34 \\
    KOI252.01 & 2.64 & 0.15 & 0.111 & 0.001 & 362 & 10 & -196 & 974 & 142.4 & -2.84 & 9.70 \\
    KOI253.01 & 2.73 & 0.15 & 0.057 & 0.004 & 530 & 7 & -613 & 891 & 176.1 & -3.62 & 5.54 \\
    Kepler-505b & 2.60 & 0.10 & 0.150 & 0.001 & 314 & 20 & 896 & 360 & 96.5 & 8.67 & 4.04 \\ 
    Kepler-138c & 1.20 & 0.07 & 0.094 & 0.001 & 402 & 50 & -1561 & 2052 & 40.6 & -23.15 & 36.06 \\
    Kepler-138d & 1.21 & 0.08 & 0.133 & 0.001 & 338 & 20 & -391 & 1352 & 141.9 & -5.11 & 26.74 \\
    Kepler-205c & 1.58 & 0.09 & 0.126 & 0.001 & 389 & 50 & 196 & 1761 & 92.7 & 2.45 & 23.34 \\
    Kepler-236c & 2.00 & 0.10 & 0.136 & 0.001 & 329 & 5 & 718 & 2040 & 131.4 & 9.68 & 25.67 \\
    Kepler-249d & 1.08 & 0.10 & 0.095 & 0.001 & 332 & 5 & -523 & 1900 & 138.2 & -9.53 & 44.18 \\
    Kepler-737b & 1.96 & 0.11 & 0.151 & 0.001 & 298 & 50 & -494 & 781 & 132.3 & -7.97 & 11.50 \\
    Kepler-32d & 2.50 & 0.11 & 0.130 & 0.001 & 328 & 15 & 15 & 790 & 71.0 & 0.01 & 7.79 \\
    Kepler-36b & 2.24 & 0.10 & 0.261 & 0.001 & 273 & 13 & 1053 & 1426 & 130.2 & 15.91 & 18.70 \\
    Kepler-395c & 1.34 & 0.07 & 0.177 & 0.001 & 341 & 27 & 4739 & 1817 & 58.7 & 74.34 & 49.03 \\
    K2-3b & 2.18 & 0.30 & 0.076 & 0.002 & 463 & 40 & 482 & 299 & 148.9 & 3.74 & 2.66 \\
    K2-3c & 1.85 & 0.27 & 0.138 & 0.001 & 311 & 30 & -215 & 538 & 88.9 & -2.46 & 8.11 \\
    K2-3d & 1.51 & 0.23 & 0.205 & 0.001 & 255 & 30 & -122 & 1019 & 30.5 & -1.79 & 21.11 \\
    K2-9b & 2.25 & 0.96 & 0.098 & 0.001 & 391 & 70 & -494 & 252 & 73.3 & -4.35 & 2.87 \\
    K2-18b & 2.38 & 0.22 & 0.150 & 0.001 & 266 & 15 & 232 & 308 & 81.4 & 3.03 & 4.12 \\
    K2-21b & 1.84 & 0.10 & 0.078 & 0.002 & 480 & 20 & 5228 & 1142 & 95.7 & 46.09 & 17.76 \\
    K2-21c & 2.49 & 0.17 & 0.110 & 0.001 & 405 & 20 & 2315 & 808 & 96.3 & 19.41 & 8.17 \\
    K2-26b & 2.58 & 0.20 & 0.094 & 0.007 & 409 & 20 & 814 & 395 & 121.1 & 6.5 & 3.5 \\
    K2-28b & 2.32 & 0.24 & 0.020 & 0.049 & 568 & 35 & -75 & 40 & 201.8 & -0.45 & 0.27 \\
    EPIC 210558622 & 3.20 & 1.80 & 0.082 & 0.013 & 381 & 50 & -7 & 69 & 140.5 & -0.07 & 0.50 \\
    \hline
    GJ 1132b & 1.43 & 0.16 & 0.015 & 0.016 & 644 & 38 & -277 & 122 & 76.1 & -1.93 & 1.32 \\
    GJ 1214b & 2.85 & 0.20 & 0.014 & 0.006 & 604 & 19 & -40 & 47 & 195.7 & -0.19 & 0.25 \\
    \hline
    \end{tabularx}
    \label{tab:slope}
\bigskip
\end{table*}

\subsection{Two Populations of Planets}\label{sec:2population}

\begin{figure*}
\begin{center}
    \subfigure
	{%
	\label{fig:1Gauss}
	\includegraphics[width=0.45\textwidth]{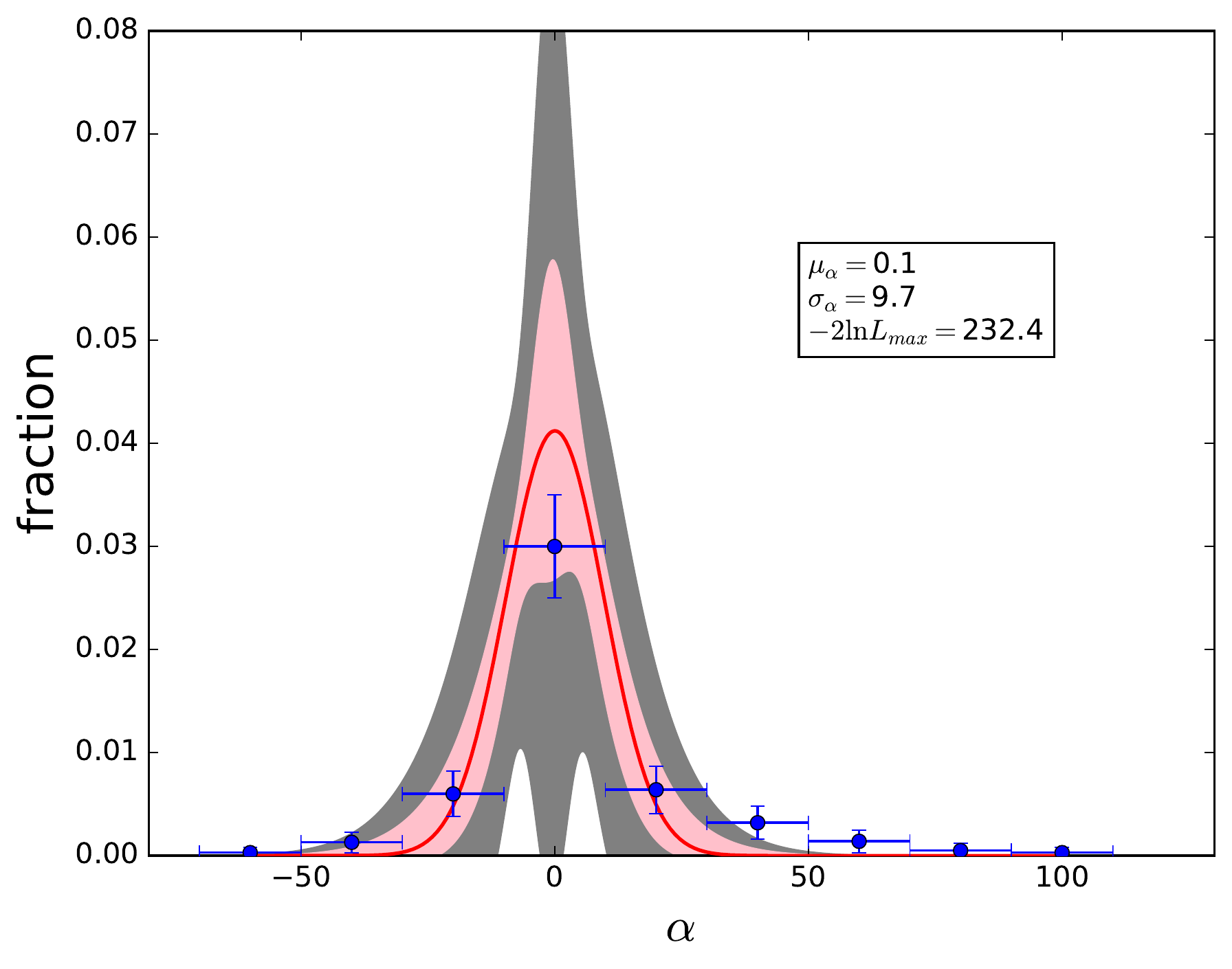}
	}%
    \subfigure
	{%
	\label{fig:2Gauss}
	\includegraphics[width=0.45\textwidth]{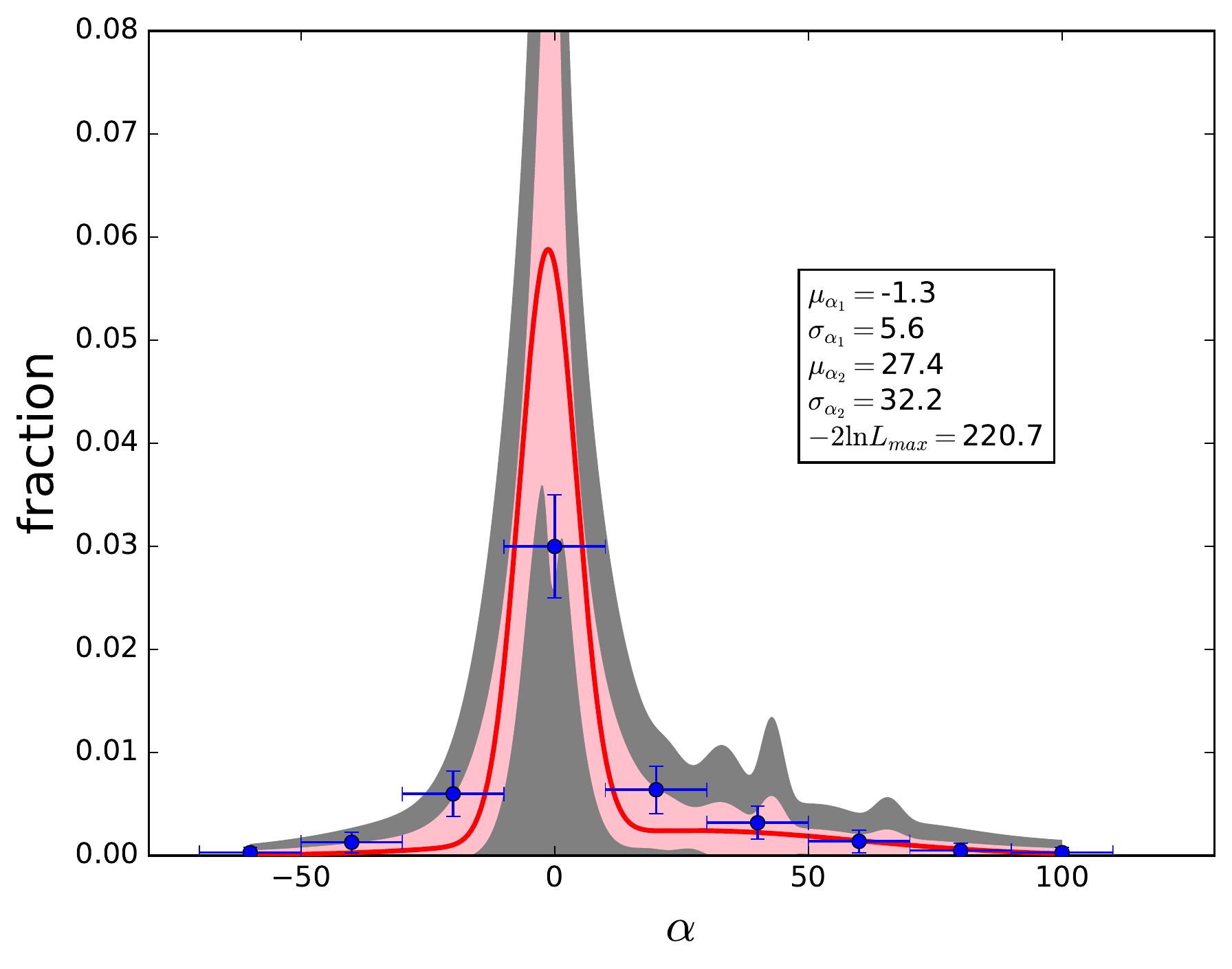}
	}
\end{center}
\caption{(a) Best-fit distribution for the 1-Gauss model (in red), in comparison with the binned distribution of $\alpha$ values of planets in our sample (blue points). These binned blue points are only presented for demonstrations, and the actual fitting does not depend on the binning. The pink and grey shaded areas represent the $1\sigma$ and $3\sigma$ posterior distributions of the fitting. The center of the best-fit Gaussian is $\mu_{\alpha}=0.1$, and standard deviation is $\sigma_{alpha}=9.7$. (b) Same as (a), but for the 2-Gauss model. The center and standard deviation of the smaller population are $\mu_{\alpha 1} = 27.4$ and $\sigma_{\alpha 1} = 32.2$ respectively, and the center and standard deviation of the main population are $\mu_{\alpha 1} = -1.3$ and $\sigma_{\alpha 1} = 5.6$ respectively. The fraction of the smaller population is $20\pm 10\%$ in the best-fit model. As is discussed in section \ref{sec:2population}, the 2-Gaussian model yields a much better fit than the 1-Gaussian model, indicating a high probability that the kind of planets in our sample need to be described as 2 populations.}
\label{fig:GaussModel}
\bigskip
\end{figure*}

We fit the distribution of $\alpha$ values of the 30  planets with two models: one with a single Gaussian shape (1-Gaussian model), and the other with two Gaussian formulae added together (2-Gaussian model). 
The 2-Gaussian model is defined by the separate width $\sigma$ and the mean $\mu$ of each Gaussian, in addition to a fraction parameter $f$ describing the fraction of planets going into the population modeled by the second Gaussian, so that we allow $\sigma_{\alpha 1}$, $\sigma_{\alpha 2}$, $\mu_{\alpha 1}$, $\mu_{\alpha 2}$ and $f$, to vary. 

When $f=0$, the 2-Gaussian model collapses to the 1-Gaussian model. Our set of models is therefore “nested,” because all instances of the simpler model are a subset of the more complex model. The goodness of fit of the two models are compared using the Likelihood Ratio Test \citep{Likelihood1997}, which is applicable for nested models. The test statistics are defined as:

\begin{equation}\label{Likelihood}
\begin{aligned}
D = & -2\times [\ln{\rm (likelihood~for~1~Gaussian~model)}\\
&- \ln{\rm (likelihood~for~2~Gaussian~mode})],
\end{aligned}
\end{equation}

\noindent
and the probability distribution of $D$ is approximately a $\chi$-squared distribution with degrees of freedom equal to the number of new parameters introduced by the more complicated model, which in our case is 3. Therefore by calculating $D$ and comparing with the relevant $\chi$-squared distribution, we are able to obtain the $p$-value of the more complicated 2-Gaussian model when comparing with the 1-Gaussian model, which indicates the probability that 2 populations are needed to fit the observed $\alpha$ distribution.

Figure \ref{fig:1Gauss} summarizes the results of this exercise. For the sake of illustration, we depict the underlying $\alpha$ distribution in blue. For each planet, we generate a random set of $\alpha$ values drawn from the mean and uncertainty of the measured $\alpha$. We then combine these sample distributions for all planets. The result of binning this final, combined distribution is shown in blue. These binned blue points are only presented for demonstrations, and the actual fitting does not depend on the binning. The red curve is the best-fit model after maximizing the likelihood, which shows a distribution of $\alpha$ centered on $\mu_{\alpha} = 0.1$ with a standard deviation of $\sigma_{\alpha} = 9.7$. The pink and grey shaded areas represent the $1\sigma$ and $3\sigma$ posterior distributions of the fitting. A visual comparison between the likeliest model (red) and the underlying distribution (blue) shows that a single Gaussian does not adequately recover the tail of high $\alpha$ values.

In comparison, the 2-Gaussian model proves to be a significantly better fit for the observed data, as is shown in Figure \ref{fig:2Gauss}. The minor population with high $\alpha$ values is represented by a Gaussian centered at $\mu_{\alpha_1} = 27.4$ with standard deviation $\sigma_{\alpha_1} = 32.2$, and the major population with lower $\alpha$ is represented by a Gaussian centered at $\mu_{\alpha_2} = -1.3$ with standard deviation $\sigma_{\alpha_2} = 5.6$. The fraction of the smaller population is $20\pm 10\%$ in the best-fit model. And for this model, $-2\ln{\rm (maximum~likelihood)}\approx 220.7$.

From above, we calculate the test statistics $D$ to be 11.7, and on the $\chi$-square distribution with degree of freedom 3, this corresponds to a $p$-value smaller than 0.01.
We conclude that the $\alpha$ values from our sample of planets are not drawn from a single Gaussian distribution, but rather from two distinct Gaussian distributions. The data favor the latter hypothesis with odds of 100:1.

To investigate the possibility of a third population, we also fit the $\alpha$ distribution with a 3-Gaussian model, and found $-2\ln{\rm (maximum~likelihood)}\approx 220.7$ for this model, which corresponds to a $p$-value larger than 0.95 when comparing the 2-Gaussian and the 3-Gaussian model. The result indicates that a 2-Gaussian model is sufficient to describe the $\alpha$ distribution with a probability larger than 95:100.

\bigskip

\section{Discussion}\label{sec:discussion}

\cite{Horst2018} presented the first laboratory haze production experiments for atmospheres of super-Earths and mini-Neptunes. They explored a temperature range similar to that of our sample, and found that while all simulations produced particles, the production rates varied greatly, with some rates as low as 0.04 mg/hour. This experimental result predicts that some, but not all super-Earth and mini-Neptune atmospheres possess photochemical hazes, which is consistent with our conclusion that the smaller population indicates clear atmospheres with detectable molecular features. More implications of our results for the atmospheric and physical properties of planets are discussed as follows.

\subsection{Atmosphere of planets around early M/late K dwarfs}

UV emissions from M~type planet host stars have extensive implications on planets' habitability. For planets orbiting close to their M~type hosts, the large amount of UV flux can drive the photochemistry processes which lead to the formation of Titan-like hazes in $\rm CH_4$-rich planetary atmospheres \citep{Turbet2017}, and thus result in negative $\alpha$ values. In addition, Extreme-UV photons are capable of ionizing hydrogen atoms, heating the atmosphere, and triggering atmosphere escaping which will present the transmission spectrum as a flat line \citep{France2012}. These theories are consistent with the trends we see in Figure \ref{fig:alpha_SMA} and \ref{fig:alpha_Tp}, that planets closer to the stars or with higher equilibrium temperatures tend to have flat or negatively sloped transmission spectra. Although the relatively large uncertainties in \spitzer\ transit depth measurements, the small number of planets with high $\alpha$ values, and the fact that we only have information on two or three bandpass of planet transmission spectra hinder us from confirming any trend with high significance, we are confident in making two points about planets around M stars with our sample: first, it is possible for temperate planets around early~M/late~K dwarfs to retain an atmosphere; and second, temperate planets around early~M/late~K dwarfs consist of 2 populations in terms of atmosphere power law index $\alpha$, as was discussed quantitatively in Section \ref{sec:2population}.

\subsection{Puffy Planets}\label{sec:mass_discussion}

Both \cite{Rogers2015} and \cite{Wolfgang2016} identified a discontinuity in planet composition at $\rm 1.6~R_{Earth}$. The majority of planets smaller than this size are rocky, though \cite{Wolfgang2016} also found larger uncertainty in the mass-radius relationship among planets this size. \cite{Weiss2014} also found that planets smaller than $1.5~R_{\rm p}$ have smaller bulk density, and therefore shallower $M_{\rm p}-R_{\rm p}$ relation with an uncertainty of $2.7~\rm M_{\oplus}$ in mass. These could explain why three planets smaller than $2~\rm R_{\oplus}$ in our sample show $\alpha \gtrapprox 20$: the $M_{\rm p}-R_{\rm p}$ relation we use could have overestimated their density, leading to surface gravity $g$ larger than true values, therefore too small scale heights and too large absolute values of the power law index $\alpha$ as a result. If accurate mass measurements can be obtained for these planets, we will be able to correct for this enlargement effect in $\alpha$ measurements.

These planets might conceivably be drawn from the lower density mass-radius relation in \cite{Wolfgang2016} for planets with masses measured by transit-timing variation. This category of planets already includes at least one planet from this sample: Kepler-138d. 

\cite{Lee2016} identified a type of planet they designated as ``super-puff'', which contains only 2-6~$\rm M_{Earth}$ of material with radii between 4--10~$\rm R_{Earth}$. These super-puffs acquired their thick atmospheres as dust-free rapidly cooling worlds outside $\sim$1~AU. Although our planets are smaller and closer to their host stars than the super-puffs described in \cite{Lee2016}, their trend in temperature, as is shown in Figure \ref{fig:alpha_Tp}, matches the hypothesis that these puffy planets formed in cooler environment.

The super-puff hypothesis will not affect our argument of 2 populations of planets. As can be observed in Figure \ref{fig:alpha_Rp}, the trend that smaller planets may be puffier only exists for a subset of our planet sample, while the rest of our planets show a consistent $\alpha$ distribution around zero slope over all sizes.


\section{Summary and Conclusion}

We analyzed the \spitzer\ light curves of 28 temperate super-Earths/mini-Neptunes orbiting early M dwarfs or late K dwarfs using the PLD method combined with a Gaussian process and obtained their transit depths with uncertainties on the \spitzer\ 4.5~$\mu$m band. These planets also have previous transit depth measurements in the \kepler\ bandpass, so that we can study the large-scale slope of their transmission spectra. While we cannot draw definitive atmospheric conclusions from transits in only two wavelength regions, the difference between optical and infrared transit depth is informative: first, a significant difference of any kind necessitates the presence of an atmosphere. And secondly, the difference in transit depth between these two wavelength regions is correlated with strength of molecular absorption in Jovian planets \citep{Sing2016}.

Adding two other previously studied planets with similar physical and orbital properties to our sample, we found that the distribution of the index $\alpha$ of these 30 small planets cannot be described by a single Gaussian model. Quantitatively, we compared a 1-Gaussian model fitting of the distribution with a 2-Gaussian model fitting, which introduces 3 extra parameters, using the Likelihood Ratio Test, and found a $p$-value smaller than 0.01, indicating a high probability that a 2 population model better fits the distribution of the index $\alpha$. The main population, consisting of $80\pm 10\%$ of the planets as was shown in our fitting, has a mean $\alpha$ of -1.3 with a standard deviation of 5.6, indicating atmospheres with some level of hazes or clouds, or bare planets with atmospheres evaporated. The smaller population, consisting of around $20\pm10\%$ planets in our sample, show high $\alpha$ values with a mean of 27.4 and a standard deviation of 32.2, which hints relatively clear atmospheres with detectable molecular features on the IR region.

We also plotted $\alpha$ values against various planet properties. The smaller population which favors high $\alpha$ values seems to consist of smaller planets, namely super-Earths with radius $\lessapprox 2 R_{\oplus}$, while the main population is distributed similarly across all $R_p$ values. Although no obvious correlation is found between planet semi-major axis and their $\alpha$ values, a relatively prominent trend is observed on the $\alpha-T_{\rm p}$ plot, which shows that all planets with larger power-law slope $\alpha$ are located in the region with $T_{\rm p}<500$~K. 

As discussed in Section \ref{sec:mass_discussion}, the measurements of planet masses are critical in calculating their atmosphere scale heights and decoding the atmosphere compositions. Therefore RV measurements and TTV analysis of higher precision are needed to characterize temperate super-Earths/mini-Neptunes.

On the other hand, we have shown that some of the temperate super-Earth/mini-Neptune sized planets orbiting around early M/late K dwarfs can retain atmosphere with detectable molecular features, and it is worth devoting \HST/\JWST\ time to observe selected planets of this kind. In particular, we are interested in eight planets in our sample that have $\alpha$ values inconsistent with zero: K2-28b, K2-26b, K2-3b, K2-21b, K2-21c, KOI247.01, Kepler-505b, Kepler-395c and Kepler-125b. We will assess the feasibility to observe them in wider optical/IR bandpass and evaluate their transmission spectra in future works to prepare for the upcoming $\it JWST$ mission.

\bigskip

\section{Acknowledgements}

This research has made use of the Exoplanet Follow-up Observation Program website, which is operated by the California Institute of Technology, under contract with the National Aeronautics and Space Administration under the Exoplanet Exploration Program.

This work is based [in part] on observations made with the \spitzer\ Space Telescope, which is operated by the Jet Propulsion Laboratory, California Institute of Technology under a contract with NASA. Support for this work was provided by NASA through an award issued by JPL/Caltech.

D. Dragomir acknowledges support provided  by NASA through Hubble Fellowship grant HST-HF2-51372.001-A awarded by the Space Telescope Science Institute, which is operated by the Association of Universities for Research in Astronomy, Inc., for NASA, under contract NAS5-26555.

We thank Professor Ian J. M. Crossfield and Mr. David Anthony Berardo for helpful discussion and comments on this work.

\bibliographystyle{apj}
\bibliography{mybib}

\bigskip

\appendix

\LongTables
\renewcommand*{\arraystretch}{1.7}
\begin{longtable}{ccccp{2.5in}}
\caption{Pixels we use from each \spitzer\ transit, and the best-fit mid-transit epochs.}\label{append:pixels} \\

\hline 
\multicolumn{1}{c}{\textbf{Name}} & \multicolumn{1}{c}{\textbf{Channel}} & \multicolumn{1}{c}{\textbf{AOR}} & \multicolumn{1}{c}{\textbf{$T_0$ (BJD-2400000.5)}} & \multicolumn{1}{c}{\textbf{Image pixel coordinates (X, Y)}} \\ \hline 
\endfirsthead

\multicolumn{3}{c}%
{{\bfseries \tablename\ \thetable{} -- continued from previous page}} \\
\hline \multicolumn{1}{c}{\textbf{Name}} & \multicolumn{1}{c}{\textbf{Channel}} & \multicolumn{1}{c}{\textbf{AOR}} & \multicolumn{1}{c}{\textbf{$T_0$ (BJD-2400000.5)}} & \multicolumn{1}{c}{\textbf{Image pixel coordinates (X, Y)}} \\ \hline 
\endhead

\hline \multicolumn{3}{r}{{Continued on next page}} \\ \hline
\endfoot

\hline \hline
\endlastfoot
    KOI247.01 & 4.5 $\mu$m & 39368448 & $55400.4629\pm 0.0028$ & (127, 128--130) (128, 128--130) (129, 128--130) \\
    KOI247.01 & 4.5 $\mu$m & 39368704 & $55358.9763\pm 0.0011$ & (127, 128--130) (128, 128--130) (129, 128--130) \\
    KOI247.01 & 4.5 $\mu$m & 41164032 & $55552.3436\pm 0.0012$ & (127, 128--130) (128, 128--130) (129, 128--130) \\
    Kepler-49b & 4.5 $\mu$m & 39370496 & $55370.1832\pm 0.0126$ & (127, 128--130) (128, 128--130) (129, 128--130) \\
    Kepler-49b & 4.5 $\mu$m & 41165056 & $55557.4626\pm 0.0047$ & (127, 128--130) (128, 128--130) (129, 128--130) \\
    Kepler-49c & 4.5 $\mu$m & 39366656 & $55547.9947\pm 0.0231$ & (127, 128--130) (128, 128--130) (129, 128--130) \\
    Kepler-49c & 4.5 $\mu$m & 39366912 & $55417.0215\pm 0.0059$ & (126, 128--129) (127, 127--130) (128, 127--130) (129, 128--129) \\
    Kepler-49c & 4.5 $\mu$m & 39367168 & $55373.3786\pm 0.0030$ & (127, 128--130) (128, 128--130) (129, 128--130) \\
    Kepler-504b & 4.5 $\mu$m & 39419648 & $55390.2429\pm 0.0038$ & (127, 128--130) (128, 128--130) (129, 128--130) \\
    Kepler-504b & 4.5 $\mu$m & 39421952 & $55437.9400\pm 0.0073$ & (127, 128--130) (128, 128--130) (129, 128--130) \\
    Kepler-26c & 4.5 $\mu$m & 41196544 & $55551.7473\pm 0.0057$ & (127, 128--130) (128, 128--130) (129, 128--130) \\
    Kepler-26c & 4.5 $\mu$m & 41196800 & $55517.1879\pm 0.0069$ & (127, 128--130) (128, 128--130) (129, 128--130) \\
    Kepler-26c & 4.5 $\mu$m & 41197056 & $55499.9264\pm 0.0153$ & (127, 128--130) (128, 128--130) (129, 128--130) \\
    Kepler-125b & 4.5 $\mu$m & 39437824 & $55399.2018\pm 0.0058$ & (25, 138--140) (26, 138--140) (27, 138--140) \\
    Kepler-125b & 4.5 $\mu$m & 41164800 & $55544.9614\pm 0.0030$ & (200, 56--58) (201, 56--58) (202, 56--58) \\
    KOI252.01 & 4.5 $\mu$m & 39421696 & $55372.6851\pm 0.0017$ & (127, 128--130) (128, 128--130) (129, 128--130) \\
    KOI252.01 & 4.5 $\mu$m & 41166336 & $55566.3704\pm 0.0319$ & (127, 128--130) (128, 128--130) (129, 128--130) \\
    KOI253.01 & 4.5 $\mu$m & 41440256 & $55571.1547\pm 0.0018$ & (127, 128--130) (128, 128--130) (129, 128--130) \\
    Kepler-505b & 4.5 $\mu$m & 39420416 & $55407.6231\pm 0.0012$ & (127, 128--130) (128, 128--130) (129, 128--130) \\ 
    Kepler-138c & 4.5 $\mu$m & 44144384 & $55795.8892\pm 0.0107$ & (128, 127--129) (129, 127--129) (130, 127--129) \\
    Kepler-138d & 4.5 $\mu$m & 53901056 & $57404.7769\pm 0.0016$ & (14, 16--17) (15, 15--17) (16, 14--17) (17, 15--17) (18, 16--17) \\
    Kepler-138d & 4.5 $\mu$m & 53901568 & $57427.8772\pm 0.0046$ & (14, 16--17) (15, 15--17) (16, 14--17) (17, 15--17) (18, 16--17) \\
    Kepler-138d & 4.5 $\mu$m & 53901824 & $57312.4018\pm 0.0013$ & (14, 16--17) (15, 15--18) (16, 15--18) (17, 15--18) (18, 16--17) \\
    Kepler-138d & 4.5 $\mu$m & 53902080 & $57289.3328\pm 0.0056$ & (14, 16--17) (15, 14--18) (16, 14--18) (17, 14--18) (18, 16--17) \\
    Kepler-205c & 4.5 $\mu$m & 44158720 & $55878.3423\pm 0.0104$ & (126, 128--129) (127, 127--130) (128, 127--130) (129, 127--128) \\
    Kepler-205c & 4.5 $\mu$m & 44159232 & $55858.0391\pm 0.0050$ & (126, 128--129) (127, 127--130) (128, 127--130) (129, 127--128) \\
    Kepler-236c & 4.5 $\mu$m & 44160256 & $55857.5281\pm 0.0116$ & (127, 128--130) (128, 128--130) (129, 128--130) \\
    Kepler-236c & 4.5 $\mu$m & 44160512 & $55809.7059\pm 0.0022$ & (127, 128--130) (128, 128--130) (129, 128--130) \\
    Kepler-249d & 4.5 $\mu$m & 44165376 & $55932.7258\pm 0.0036$ & (127, 129--131) (128, 129--131) (129, 129--131) \\
    Kepler-249d & 4.5 $\mu$m & 44165632 & $55886.6196\pm 0.0138$ & (127, 128--130) (128, 128--130) (129, 128--130) \\
    Kepler-737b & 4.5 $\mu$m & 44164608 & $55880.4110\pm 0.0022$ & (127, 128--130) (128, 128--130) (129, 128--130) \\
    Kepler-737b & 4.5 $\mu$m & 44165120 & $55851.8987\pm 0.0335$ & (127, 128--130) (128, 127--131) (129, 127--131) (130, 128--130) \\
    Kepler-32d & 4.5 $\mu$m & 44159744 & $55853.3585\pm 0.0110$ & (127, 127--129) (128, 127--129) (129, 127--129) \\
    Kepler-32d & 4.5 $\mu$m & 44160000 & $55830.5824\pm 0.0060$ & (127, 127--130) (128, 127--130) (129, 127--130) \\
    Kepler-36b & 4.5 $\mu$m & 44161280 & $55822.0096\pm 0.0064$ & (127, 127--130) (128, 127--130) (129, 127--130) \\
    Kepler-395c & 4.5 $\mu$m & 45540096 & $56201.3472\pm 0.0059$ & (23, 231--233) (24, 231--233) (25, 231--233) \\
    Kepler-395c & 4.5 $\mu$m & 45540608 & $56166.3604\pm 0.0093$ & (23, 231--233) (24, 231--233) (25, 231--233) \\
    Kepler-395c & 4.5 $\mu$m & 45540864 & $56131.4289\pm 0.0116$ & (23, 231--233) (24, 231--233) (25, 231--233) \\
    K2-3b & 4.5 $\mu$m & 53521664 & $57104.4949\pm 0.0002$ & (15, 15--17) (16, 14--17) (17, 15--17) (18, 16--17) \\
    K2-3b & 4.5 $\mu$m & 53521920 & $57094.4479\pm 0.0011$ & (15, 15--17) (16, 14--17) (17, 15--17) (18, 16--17) \\
    K2-3b & 4.5 $\mu$m & 56417024 & $57275.4277\pm 0.0006$ & (14, 16--17) (15, 15--18) (16, 14--18) (17, 15--18) (18, 16--18) \\
    K2-3b & 4.5 $\mu$m & 58585856 & $57496.6253\pm 0.0012$ & (14, 17--18) (15, 16--19) (16, 15--19) (17, 16--19) (18, 17--19) \\
    K2-3b & 4.5 $\mu$m & 58587136 & $57466.4720\pm 0.0012$ & (14, 17--18) (15, 16--19) (16, 15--19) (17, 16--19) (18, 17--19) \\
    K2-3b & 4.5 $\mu$m & 59934464 & $57627.3438\pm 0.0008$ & (15, 13--16) (16, 13--16) (17, 13--16) (18, 14--16) \\
    K2-3c & 4.5 $\mu$m & 53522944 & $57107.5340\pm 0.0019$ & (14, 16--17) (15, 14--17) (16, 14--17) (17, 14--17) (18, 15--17) \\
    K2-3c & 4.5 $\mu$m & 56418816 & $57280.0624\pm 0.0019$ & (14, 16--17) (15, 15--18) (16, 14--18) (17, 15--18) (18, 16--17) \\
    K2-3c & 4.5 $\mu$m & 58235392 & $57477.2428\pm 0.0017$ & (14, 16--17) (15, 14--17) (16, 14--18) (17, 14--17) (18, 16--17) \\
    K2-3d & 4.5 $\mu$m & 53522432 & $57093.0654\pm 0.0013$ & (14, 16) (15, 15--17) (16, 14--18) (17, 14--17) (18, 15--17) \\
    K2-3d & 4.5 $\mu$m & 56419072 & $57271.2888\pm 0.0021$ & (14, 16--18) (15, 15--18) (16, 14--18) (17, 15--18) (18, 16--18) \\
    K2-9b & 4.5 $\mu$m & 56418560 & $57283.4288\pm 0.0035$ & (22, 232) (23, 231--233) (24, 230--234) (25, 230--234) (26, 231--233) \\
    K2-9b & 4.5 $\mu$m & 58586880 & $57486.3896\pm 0.0038$ & (22, 232--233) (23, 230--234) (24, 230--234) (25, 231--234) (26, 232--233) \\
    K2-18b & 4.5 $\mu$m & 56416000 & $57263.8870\pm 0.0003$ & (14, 16--17) (15, 15--18) (16, 14--18) (17, 15--18) (18, 16--17) \\
    K2-21b & 4.5 $\mu$m & 58530560 & $57427.2443\pm 0.0007$ & (12, 15--16) (13, 14--17) (14, 13--17) (15, 14--17) (16, 15--17) \\
    K2-21c & 4.5 $\mu$m & 58531072 & $57421.9558\pm 0.0003$ & (12, 15) (13, 13--16) (14, 13--17) (15, 13--17) (16, 14--16) \\
    K2-26b & 4.5 $\mu$m & 54341376 & $57168.0052\pm 0.0010$ & (22, 232--233) (23, 231--233) (24, 230--234) (25, 231--233) (26, 232--233) \\
    K2-26b & 4.5 $\mu$m & 58396672 & $57401.1104\pm 0.0018$ & (23, 231--234) (24, 230--234) (25, 231--234) (26, 232--233) \\
    K2-28b & 4.5 $\mu$m & 62339840 & $57795.7682\pm 0.0002$ & (14, 16--17) (15, 14--18) (16, 14--18) (17, 15--17) (18, 16--17) \\
    EPIC 210558622 & 4.5 $\mu$m & 57764352 & $57330.8128\pm 0.0005$ & (22, 232--233) (23, 231--234) (24, 230--234) (25, 231--234) (26, 232--233) \\
    EPIC 210558622 & 4.5 $\mu$m & 58797056 & $57554.5017\pm 0.0011$ & (22, 232--233) (23, 231--234) (24, 230--234) (25, 231--234) (26, 232--233) \\
    \hline
    Kepler-138d & 3.6 $\mu$m & 53910528 & $57381.6743\pm 0.0231$ & (14, 16--17) (15, 14--17) (16, 14--17) (17, 15--17) (18, 16--17) \\
    Kepler-138d & 3.6 $\mu$m & 53910784 & $57335.4944\pm 0.0025$ & (15, 15--18) (16, 15--18) (17, 15--18) (18, 16--18) \\
    Kepler-138d & 3.6 $\mu$m & 53911040 & $57266.2288\pm 0.0115$ & (14, 16--17) (15, 15--17) (16, 14--17) (17, 15--17) (18, 16--17) \\
    Kepler-138d & 3.6 $\mu$m & 53911296 & $57243.0202\pm 0.0012$ & (14, 16--17) (15, 15--17) (16, 14--17) (17, 15--17) (18, 16--17) \\
    K2-18b & 3.6 $\mu$m & 58234368 & $57461.5438\pm 0.0004$ & (14, 16--17) (15, 14--17) (16, 14--17) (17, 14--17) (18, 16--17) \\

\end{longtable}

\end{document}